\documentclass[11pt,a4paper]{article}
\pdfoutput=1
\usepackage{jheppub}
\usepackage{epsfig}
\usepackage{graphicx}
\usepackage{subfigure}
\usepackage{dcolumn}
\usepackage{bm}
\usepackage{float}
\usepackage{cancel}

\def\beq{\begin{equation}}
\def\eeq{\end{equation}}
\def\bea{\begin{eqnarray}}
\def\eea{\end{eqnarray}}
\def\to{\rightarrow}

\title{Impact of LHC Searches on NLSP Top Squark and Gluino Mass}

\author[a]{Bin He}
\author[a,1]{Tong Li\note{corresponding author}}
\author[a]{Qaisar Shafi}

\affiliation[a]{Bartol Research Institute, Department of Physics and Astronomy,\\
University of Delaware, Newark, Delaware 19716, USA}

\emailAdd{hebin@udel.edu}
\emailAdd{tli@udel.edu}
\emailAdd{shafi@bartol.udel.edu}

\arxivnumber{1112.4461}

\abstract{

We explore the implications of 7 TeV LHC searches for a scenario in which one of the stops is the next-to-lightest supersymmetric particle (NLSP). The NLSP stop ($\tilde{t}_1$) is assumed to decay exclusively into neutralino and charm quark. We consider processes where the stops are pair produced together with a hard QCD jet. We also consider stop quarks from gluino decays, $\tilde{g}\to t\tilde{t}_1^\ast+\bar{t}\tilde{t}_1$. We show that the monojet ATLAS and CMS searches corresponding to 1 fb$^{-1}$ of integrated luminosity are sensitive to stop masses of up to 160 GeV, with the 20\% neutralino-stop coannihilation region  essentially ruled out for $M_{\tilde{t}_1}\lesssim 140$ GeV. The region $M_{\tilde{t}_1}\lesssim 130$ GeV is excluded with even relatively larger mass difference, $M_{\tilde{t}_1}-M_{\tilde{\chi}_1^0}\sim 40$ GeV, by the multi-jets search. The b-jet and same-sign dilepton searches are sensitive to a heavier gluino because they only pick up gluino pair production events followed by top quarks decaying into b-jets and same-sign dileptons, respectively. We find that the LHC data places a lower limit on the gluino mass in this scenario of about 600 GeV (700 GeV) from b-jets (same-sign dileptons) searches.
}


\begin{document}
\maketitle

\section{Introduction}
Low scale supersymmetry (SUSY), augmented by an unbroken R-parity, largely overcomes the gauge hierarchy problem encountered in the Standard Model (SM)~\cite{Arnowitt:2006bb} and also provides a compelling cold dark matter candidate~\cite{DM}. In a recent paper, hereafter called~\cite{ournlsp}, we explored the implications of the recent ATLAS and CMS searches for the NLSP stop scenario. The neutralino-stop coannihilation scenario can arise in realistic supersymmetric SU(5) and SO(10) models with $b-\tau$ Yukawa unification at $M_{GUT}$~\cite{btau}.
In Ref.~\cite{ournlsp} constraints on the NLSP stop mass were derived using the models presented in Ref.~\cite{btau}. Among other things, this analyses essentially ruled out models in which the NLSP stop mass lies below around 140-160 GeV, with the stop-neutralino mass difference of around 20\% or less. Other recent papers related to the NLSP stop scenario are in Refs.~\cite{wacker1,kats,sundrum,papucci,bi,in} (other related scenarios include bino-sbottom coannihilation~\cite{nlspsbottom}).

Motivated by these considerations in this paper we pursue a model-independent analysis of the NLSP stop scenario.
The search for NLSP stop, especially in the region of nearly degenerate stop and LSP neutralino masses, is challenging and has been implemented by both LEP and Tevatron~\cite{pdg,jose,cdf,cdf2}, assuming the loop-induced NLSP stop two-body decay into a charm quark and a neutralino is dominant~\cite{kobayashi,drees,drees1,shih}. The NLSP stop mass limit is $M_{\tilde{t}_1}>100$ GeV from LEP-II and $M_{\tilde{t}_1}>180$ GeV from CDF Run-II. However, the Tevatron is not sensitive to stop searches if the stop and LSP neutralino mass difference is below 40 GeV. Thus the Tevatron bound does not cover the coannihilation region above the LEP limit.

Two alternative search methods have been pointed out and implemented to detect a light stop instead of searching for events containing two jets and missing transverse energy. One of them takes advantage of the Majorana fermion feature of the gluino and considers gluino pair production followed by gluino decay into an on-shell stop and top quark~\cite{kraml,martinss}. The pair production of gluinos leads to events containing a pair of same-sign top quarks plus two same-sign stops. The benefit of this search is the anomalous same-sign dileptons signature arising from the same-sign top quarks leptonic decay, with negligible SM backgrounds. The other proposed method is to consider stop pair production in association with a hard QCD jet~\cite{wagnerstop}. In the coannihilation region, there will be minimal hadronic activity associated with the stop decay, and therefore this channel would effectively lead to events with a hard jet and large missing energy. This signature has been proposed to explore large extra dimensions~\cite{led}, search for relatively light gluinos at the Tevatron~\cite{jgogo} and for nearly degenerate gaugino pair production~\cite{tao}.

The ATLAS and CMS experiments have presented their analysis results for events containing (a) jets plus missing transverse momentum~\cite{atlasjets,cmsjets,cmsjets2}, (b) monojet plus large missing energy~\cite{atlasmonojet}, (c) b-jets with or without lepton plus missing energy~\cite{atlasb0l,atlasb1l}, and (d) two same-sign isolated leptons plus hadronic jets and missing energy~\cite{cmsss} in the final state, corresponding to an integrated luminosity of 1 fb$^{-1}$, an update of their 2010 data~\cite{atlas2010,cms2010}. Good agreement was observed between the number of events in data and the SM predictions. These searches are relevant for both NLSP stop search modes mentioned above, namely stop pair production with the emission of hard QCD jet(s) ((a) and (b)), and gluino pair production with the gluino decaying into top and stop ((c) and (d)). Thus, one could obtain improved constraints on the relevant parameter space, beyond what is probed by LEP and Tevatron.

To explore the NLSP stop scenario in a model-independent way, we adopt the so-called ``simplified models'' paradigm~\cite{simplified1,simplified2,simplified3,simplified4}. These models parameterize the new physics by a simple particle spectrum, its production modes and decay topologies, with the masses, cross sections and branching ratios taken as free parameters.
The particles that are not involved in a specific signature are assumed to be decoupled. For our case, we will consider both stop and gluino pair production, with 100\% gluino decay into top and stop and 100\% stop decay into charm and neutralino. In total, we have three parameters, namely the gluino mass $M_{\tilde{g}}$, the NLSP stop mass $M_{\tilde{t}_1}$, and the neutralino LSP mass $M_{\tilde{\chi}_1^0}$. The assumption is that all other superparticles decouple. The coannihilation requirement of NLSP stop and LSP neutralino can further reduce the number of parameters.

The paper is organized as follows. In section II we discuss the NLSP stop decay and production modes and outline the selection requirements employed by the LHC collaborations. The kinematic features of NLSP stop production are presented together with constrained parameter space in terms of the masses of gluino, stop and neutralino in section III. Our conclusions are summarized in section IV.

\section{NLSP Stop and LHC}

\subsection{Production and decay modes of NLSP stop}
In the framework of MSSM with gravity mediated supersymmetry breaking, the NLSP stop $\tilde{t}_1$, with LSP neutralino, has the following decay channels
\begin{eqnarray}
\tilde{t}_1\to c\tilde{\chi}_1^0, f\bar{f}'b\tilde{\chi}_1^0, bW^+\tilde{\chi}_1^0, t\tilde{\chi}_1^0.
\end{eqnarray}
Here $f$ and $f'$ stand for SM leptons or quarks. These decays are all generated at tree level except for the first channel, which is loop-induced and proceeds through off-diagonal elements of the CKM matrix. The three tree level channels gradually come into play from left to right, corresponding to increasing $\Delta M\equiv M_{\tilde{t}_1}-M_{\tilde{\chi}_1^0}$. They proceed through both off-shell top quark and $W$ boson exchange (or sbottom, sleptons, sneutrino, charginos), only off-shell top quark (or sbottom, charginos), and via on-shell top quark respectively. In the region of nearly degenerate NLSP stop and LSP neutralino masses that we are interested in, the last two tree level channels are both kinematically forbidden, so that the loop-induced NLSP stop two-body decay into a charm quark and a neutralino is generally considered to 
be the dominant decay mode~\cite{kobayashi,drees,drees1,shih,wagnerstop}. Experimentally, at a hadron collider, for a given $\Delta M$, the NLSP stop decay products from the 4-body channel (including leptons) are much softer and thus harder to detect compared to the 2-body channel, and have not been searched so far. Therefore, we focus on the parameter region of $M_{\tilde{t}_1}$ and $M_{\tilde{\chi}_1^0}$ with the 2-body decay being the unique NLSP stop decay channel $BR(\tilde{t}_1\to c\tilde{\chi}_1^0)\approx 100\%$. Also, we assume the total widths of the stops we study are sizable enough to guarantee the stops promptly decay in the detector. The decay length is too short to observe the displaced vertex.

\begin{figure}[tb]
\centering
\includegraphics[width=0.46\textwidth]{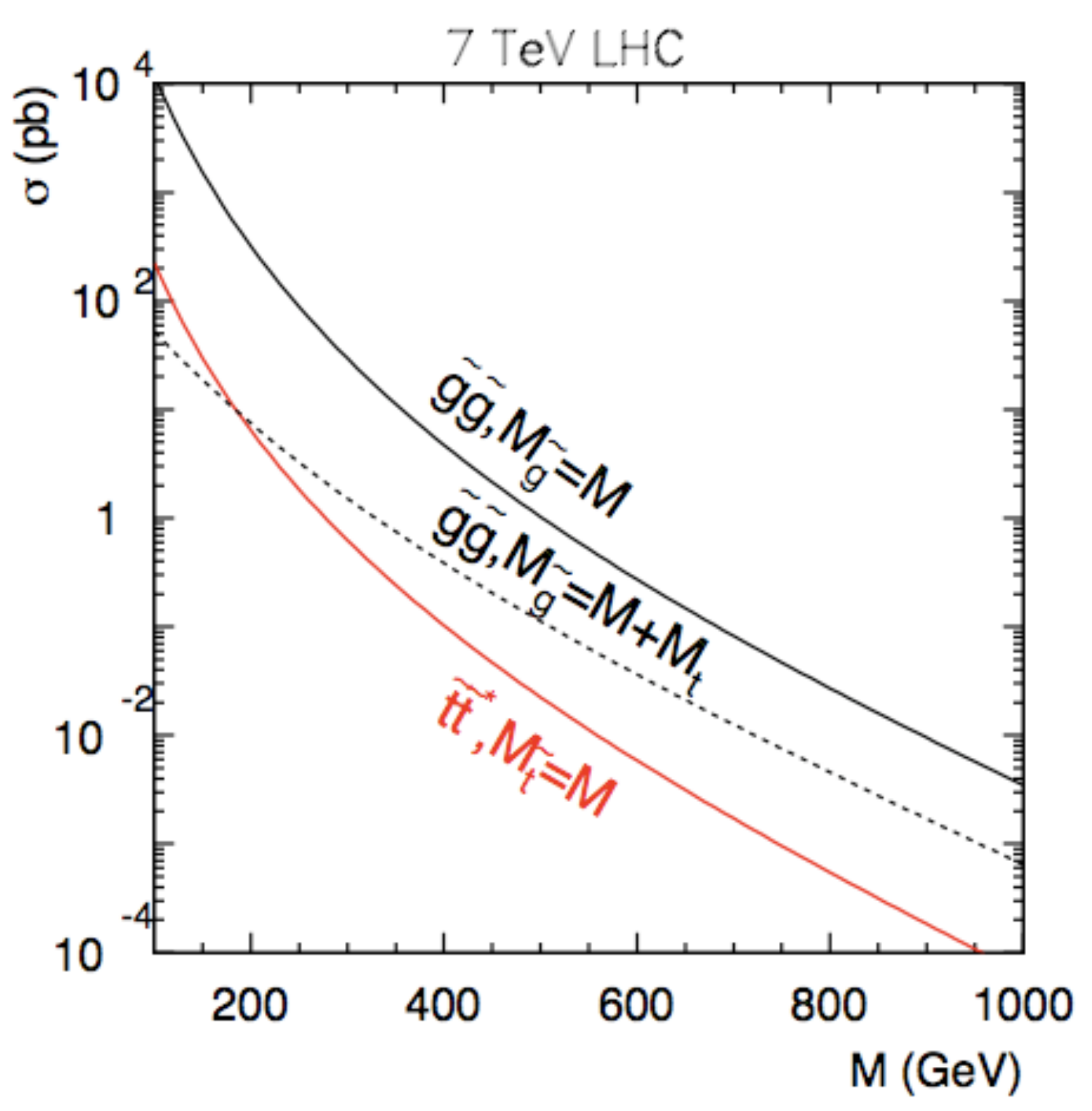}
\caption{Total cross sections for gluino and stop pair production at 7 TeV LHC. The dashed line represents gluino pair production with $M_{\tilde{g}}=M_{\tilde{t}_1}(M)+M_t$.}
\label{cs}
\end{figure}

For suitably low $M_{\tilde{t}_1}$ values, the stop pair production cross section is dominant, as shown in Fig.~\ref{cs}. However, the small mass difference between the NLSP stop and LSP neutralino means that the charm jets from the NLSP stop decay are very soft, i.e. the missing energy of these jets is very low. This scenario very likely evades the current LHC search bounds or, at best, only a tiny range of very light NLSP stop could be constrained.
It is therefore important to include the hard QCD radiation at the matrix element level in order to provide a hard jet and large missing energy. Also, in this scenario the heavier gluino essentially decays into an on-shell NLSP stop plus a top quark, $\tilde{g}\to t\tilde{t}_1^\ast+\bar{t}\tilde{t}_1$. The three-body gluino decay channels, namely $\tilde{g}\to t\bar{t}\tilde{\chi}^0, t\bar{b}\tilde{\chi}^-(\bar{t}b\tilde{\chi}^+)$, are all suppressed if the above two-body channel is open.
The energetic objects from the top decay could compensate the possibility of losing events arising from the relatively low gluino production cross section, and NLSP stop decay leading to the soft jet. More importantly, the b-jet from the top quark decay and same-sign top quarks arising from the Majorana fermion nature of the gluino provide identifiable signatures in terms of b-jet tagging and isolated same-sign dileptons in the detector.
Based on these arguments, without loss of generality, we generate hard scattering processes of gluino and NLSP stop pair production, together with the same processes with one extra jet at the matrix element level, using Madgraph/Madevent~\cite{MG}
\begin{eqnarray}
pp\to \tilde{g}\tilde{g},\ \tilde{t}_1\tilde{t}_1^\ast,\ j\tilde{t}_1\tilde{t}_1^\ast.
\end{eqnarray}
The gluino decays into $t\tilde{t}_1^\ast$ and $\bar{t}\tilde{t}_1$ with 50\% branching ratio each, and we explore the mass ranges $M_{\tilde{\chi}_1^0}+M_c<M_{\tilde{t}_1}<M_{\tilde{\chi}_1^0}+M_b+M_W$ and $M_{\tilde{g}}>M_{\tilde{t}_1}+M_t$. Also, we use Pythia to include decays, parton showering and hadronization~\cite{Pythia}, and PGS-4 to simulate the important detector effects with ATLAS/CMS-like parameters~\cite{PGS}. We must match correctly (without double-counting) between matrix element and shower generation of additional jets. In Madgraph/Madevent running, we implement MLM matching with $P_T$-ordered showers and the shower-$K_T$ scheme with $Q_{cut}=100$ GeV as described in Ref.~\cite{matching}. As a cross check, we tried various values of $Q_{cut}$ and found the uncertainty of the events generated is within $10 \%$. Therefore, $Q_{cut}=100$ GeV is used throughout our analysis. The cross sections are normalized to the next-to-leading order output of Prospino 2.1~\cite{prospino}.

\subsection{Signal selection requirements at the LHC}

The ATLAS and CMS collaborations have reported data in terms of events containing large missing transverse momentum and jets (with or without b-jets) in $\sqrt{s}=7$ TeV proton-proton collisions, corresponding to an integrated luminosity of 1 fb$^{-1}$. Also, the ATLAS experiment has searched for monojet plus missing energy events~\cite{atlasmonojet}, and the CMS collaboration has released results of same-sign dilepton signature~\cite{cmsss} with the same integrated luminosity. No excess above the SM background expectation was observed. With stricter selection cuts and more data, new upper bounds on non-SM cross sections that are at most 100 times more stringent than the 2010 results have been obtained. This data can be employed, as we show below, to find useful constraints on the NLSP stop scenario.

In the updated analysis for multi-jets and missing energy from ATLAS, the events are classified into 4 regions ``S1'', ``S2'', ``S3'' and ``S4'', where S1, S2, S3, S4 respectively requires at least 2, 3, 4, 4 jets~\cite{atlasjets}. The cut requirements are summarized in Table~\ref{cuts1}. The transverse momentum $p_T$ of a jet is defined as
 \begin{equation}
 p_T=\sqrt{p_x^2+p_y^2}.
 \end{equation}
The missing energy $\overrightarrow{\cancel{E}}$ is defined as
 \begin{equation}
 \overrightarrow{\cancel{E}}=-\sum_{i}\overrightarrow{p}_i(visible),
 \end{equation}
where the sum runs over the momenta of all visible final state particles. The missing transverse energy $\cancel{E}_T$ is defined as
 \begin{equation}
 \cancel{E}_T=\sqrt{(\cancel{E}_{x})^2+(\cancel{E}_{y})^2}.
 \end{equation}
The effective mass $m_{eff}$ is defined as the sum of $\cancel{E}_T$ and the magnitudes of the transverse momenta of the two, three or four highest $p_T$ jets used in specific signal region. $\Delta \phi(\vec{p}_T^{{\rm miss}}, j_{1,2,3})$ is the smallest azimuthal separation between the $\cancel{E}_T$ direction and the three leading jets, and $m_{eff}$ is the scalar sum of $\cancel{E}_T$ and the transverse momenta of the highest $p_T$ jets (up to two for region S1, three for region S2 and four for regions S3 and S4 respectively). The 95$\%$ C.L. upper limits on effective cross section (cross section times acceptance) for non-SM processes for signal regions S1, S2, S3, S4 are also shown in the last row of Table~\ref{cuts1}.

\begin{table}[tb]
\begin{center}
\begin{tabular}[t]{|c|c|c|c|c|}
  \hline
 & S1 & S2 & S3 & S4\\
  \hline
  Number of jets & $\geq 2$ & $\geq 3$ & $\geq 4$ & $\geq 4$\\
  \hline
 Leading jet $p_T$ (GeV) & $>130$ & $>130$ & $>130$ & $>130$\\
  \hline
 Second jet $p_T$ (GeV) & $>40$ & $>40$ & $>40$ & $>40$ \\
  \hline
 Third jet $p_T$ (GeV) & $-$ & $>40$ & $>40$ & $>40$ \\
  \hline
 Fourth jet $p_T$ (GeV) & $-$ & $-$ & $>40$ & $>40$ \\
  \hline
  $\Delta \phi(\vec{p}_T^{{\rm miss}}, j_{1,2,3})$ & $>0.4$ & $>0.4$ & $>0.4$ & $>0.4$ \\
  \hline
  $m_{eff}$ (GeV) & $>1000$ & $>1000$ & $>500$ & $>1000$\\
  \hline
  $\cancel{E}_T$ (GeV) & $>130$ & $>130$ & $>130$ & $>130$\\
  \hline
  $\cancel{E}_T/m_{eff}$ & $>0.3$ & $>0.25$  & $>0.25$  & $>0.25$ \\
  \hline
  ATLAS $\sigma_{{\rm exp}}$ (pb) & $0.022$ & $0.025$ & $0.429$ & $0.027$\\
  \hline
\end{tabular}
\end{center}
\caption{Summary of selection cuts and 95$\%$ C.L. upper limits on the effective cross section for non-SM processes for signal region S1, S2, S3 and S4 containing final states with jets and missing transverse momentum with 1 fb$^{-1}$ luminosity, following the data analyses of ATLAS~\cite{atlasjets}.}
\label{cuts1}
\end{table}

For the ATLAS search for monojet plus large missing transverse momentum, the signal events are selected according to three different cut requirements, named ``LP'', ``HP'' and ``VHP''~\cite{atlasmonojet} as shown in Table~\ref{cuts2}. The LP (HP) selection requires a jet with $p_T>120$ GeV ($p_T>250$ GeV) and $|\eta^{jet}|<2$ in the final state, and $\cancel{E}_T>120$ GeV ($\cancel{E}_T>220$ GeV). Events with a second leading jet $p_T$ above 30 GeV (60 GeV) in the region $|\eta|<4.5$ are rejected. For the HP selection, the $p_T$ of the third leading jet must be less than 30 GeV, and an additional requirement on the azimuthal separation $\Delta \phi(jet,\vec{p}_T^{miss})>0.5$ between the missing transverse momentum and the direction of the second leading jet is required. The VHP selection is defined with the same requirements as in the HP region, but with thresholds on the leading jet $p_T$ and $\cancel{E}_T$ increased up to 350 GeV and 300 GeV, respectively. The 95$\%$ C.L. upper limits on effective cross section (cross section times acceptance) for non-SM processes for the signal regions LP, HP, VHP are also shown in the last row of Table~\ref{cuts2}.

\begin{table}[tb]
\begin{center}
\begin{tabular}[t]{|c|c|c|c|}
  \hline
 & LP & HP & VHP\\
  \hline
 Leading jet $p_T$ (GeV) & $>120$ & $>250$ & $>350$\\
  \hline
 Second jet $p_T$ (GeV) & $<30$ & $<60$ & $<60$ \\
  \hline
 Third jet $p_T$ (GeV) & $-$ & $<30$ & $<30$ \\
  \hline
  $\Delta \phi(\vec{p}_T^{{\rm miss}}, j_{2})$ & $-$ & $>0.5$ & $>0.5$ \\
  \hline
  $\cancel{E}_T$ (GeV) & $>120$ & $>220$ & $>300$\\
  \hline
  ATLAS $\sigma_{{\rm exp}}$ (pb) & $1.7$ & $0.11$ & $0.035$\\
  \hline
\end{tabular}
\end{center}
\caption{Summary of selection cuts and 95$\%$ C.L. upper limits on the effective cross section for non-SM processes for signal region LP, HP and VHP containing final states with monojet and missing transverse momentum with 1 fb$^{-1}$ luminosity, following the data analyses of ATLAS~\cite{atlasmonojet}.}
\label{cuts2}
\end{table}

The selected events with b-jets and missing transverse energy from ATLAS are required to have at least one jet with $p_T>130$ GeV, at least two additional jets with $p_T>50$ GeV and $\cancel{E}_T>130$ GeV~\cite{atlasb0l,atlasb1l}. At least one jet is required to be b-tagged. Events are required to have $\cancel{E}_T/m_{eff}>0.25$, and the smallest azimuthal separation between the missing energy direction and the three leading jets, $\Delta \phi_{min}$ is required to be larger than 0.4. The signal regions are characterised by the minimal number of b-jets required in the final state and by the threshold of further selection on $m_{eff}$ as shown in Table~\ref{cuts3}: 3JA ($\geq 1$ b-jet, $m_{eff}>500$ GeV), 3JB ($\geq 1$ b-jet, $m_{eff}>700$ GeV), 3JC ($\geq 2$ b-jets, $m_{eff}>500$ GeV) and 3JD ($\geq 2$ b-jets, $m_{eff}>700$ GeV). The last column of Table~\ref{cuts3} refers to selection requirement of final states with b-jets, missing energy and one lepton, denoted by L in the following. The selected events are required to have at least four jets with $p_T>50$ GeV and $\cancel{E}_T>80$ GeV. At least one jet is required to be b-tagged. One and only one tightly selected lepton must be present, and a further selection is applied on the transverse mass of the lepton and transverse missing momentum, namely $m_T>100$ GeV. The effective mass, $m_{eff}$, is defined as the scalar sum of $\cancel{E}_T$, the transverse momenta of the four leading jets and of the lepton transverse momentum and is required to be larger than 600 GeV. The 95$\%$ C.L. upper limits on effective cross section (cross section times acceptance) for non-SM processes for the signal regions 3JA, 3JB, 3JC, 3JD and L are also shown in the last row of Table~\ref{cuts3}.

\begin{table}[tb]
\begin{center}
\begin{tabular}[t]{|c|c|c|c|c|c|}
  \hline
 & 3JA & 3JB & 3JC & 3JD & L\\
  \hline
  Number of jets with $p_T>130$ GeV& $\geq 1$ & $\geq 1$ & $\geq 1$ & $\geq 1$ & $-$\\
  \hline
   Number of jets with $p_T>50$ GeV& $\geq 2$ & $\geq 2$ & $\geq 2$ & $\geq 2$ & $\geq 4$\\
  \hline
  Number of $b$-jets & $\geq 1$ & $\geq 1$ & $\geq 2$ & $\geq 2$ & $\geq 1$\\
    \hline
  Number of leptons & 0 & 0 & 0 & 0 & 1\\
  \hline
  $\Delta \phi(\vec{p}_T^{{\rm miss}}, j_{1,2,3})$ & $>0.4$ & $>0.4$ & $>0.4$ & $>0.4$ & $-$\\
  \hline
  $m_{eff}$ (GeV) & $>500$ & $>700$ & $>500$ & $>700$ & $>600$\\
  \hline
  $\cancel{E}_T$ (GeV) & $>130$ & $>130$ & $>130$ & $>130$ & $>80$\\
  \hline
  $\cancel{E}_T/m_{eff}$ & $>0.25$ & $>0.25$  & $>0.25$  & $>0.25$ & $-$\\
  \hline
  $m_T$ (GeV) & $-$ & $-$  & $-$  & $-$ & $>100$\\
  \hline
  ATLAS $\sigma_{{\rm exp}}$ (pb) & $0.288$ & $0.061$ & $0.078$ & $0.017$ & $0.046$\\
  \hline
\end{tabular}
\end{center}
\caption{Summary of selection cuts and 95$\%$ C.L. upper limits on the effective cross section for non-SM processes for signal region 3JA, 3JB, 3JC and 3JD (L) containing final states with b-jets (plus 1 lepton) and missing transverse momentum with 830 pb$^{-1}$ (1 fb$^{-1}$) luminosity, following the data analyses of ATLAS~\cite{atlasb0l,atlasb1l}.}
\label{cuts3}
\end{table}

In the CMS analyses the events considered for search regions are all required to have two leptons with the same charge, at least two jets, and $\cancel{E}_T>30$ GeV~\cite{cmsss}.
The observed upper limits on events from new physics  are represented in Table~\ref{cuts4}. Note that in order to avoid the possibly large uncertainty from the calibration of hadronic $\tau$, in the following analyses we do not include the region T1.

We apply $\sigma\times {\rm acceptance}>\sigma_{{\rm exp}}$ as exclusion requirement for each spectrum configuration, where $\sigma$ is the relevant total cross section and the acceptance is the ratio of signal events after and before selection cuts which reflects the effects of experimental efficiency.

\begin{table}[tb]
\begin{center}
\begin{tabular}[t]{|c|c|c|c|c|c|c|c|c|}
  \hline
 & I1 & I2 & I3 & H1 & H2 & H3 & H4 & T1\\
  \hline
  $H_T$ (GeV)& $> 400$ & $> 400$ & $> 200$ & $> 400$ & $>400$ & $>200$ & $>80$ & $>400$\\
  \hline
  $\cancel{E}_T$ (GeV)& $> 120$ & $>50$ & $> 120$ & $> 120$ & $>50$ & $>120$ & $>100$ & $>120$\\
  \hline
  2 leptons $p_T>10$ GeV & &  & & $\surd$ & $\surd$ & $\surd$ & $\surd$ &\\
  $\geq 1$ lepton $p_T>20$ GeV & &  &  &  &&&&\\
   \hline
  CMS $N_{{\rm exp}}$ & $3.7$ & $8.9$ & $7.3$ & $3.0$ & $7.5$ & $5.2$ & $6.0$ & $5.8$\\
  \hline
\end{tabular}
\end{center}
\caption{Summary of selection cuts and 95$\%$ C.L. upper limits on event number for signal region I1-I3, H1-H4 and T1 containing final states with same-sign isolated dilepton, jets and missing energy with 0.98 fb$^{-1}$ luminosity, following the data analyses of CMS~\cite{cmsss}.}
\label{cuts4}
\end{table}

\section{LHC Constraints on NLSP Stop}

\subsection{NLSP stop pair production}
\begin{figure}[tb]
\centering
\includegraphics[width=0.49\textwidth]{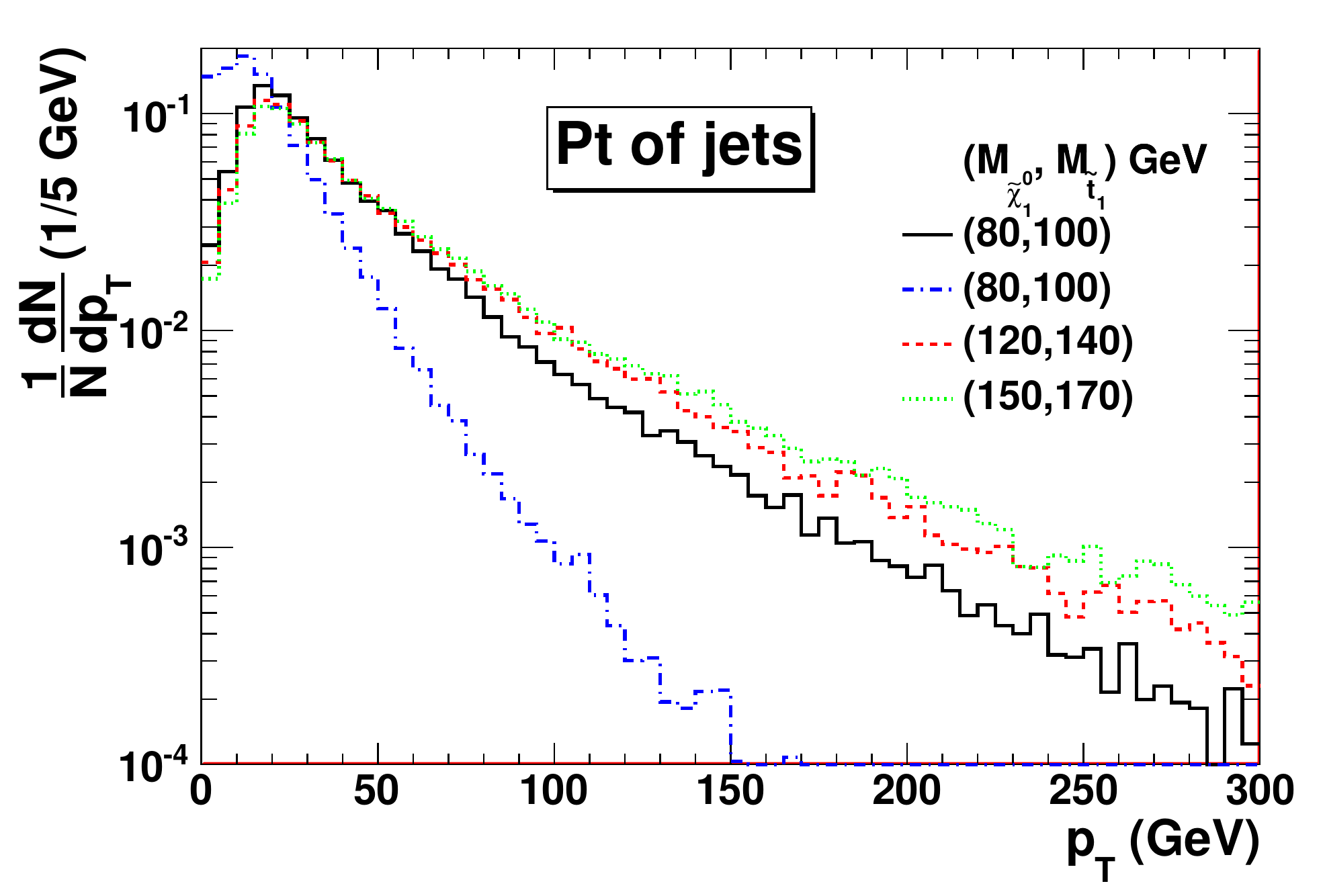}
\includegraphics[width=0.49\textwidth]{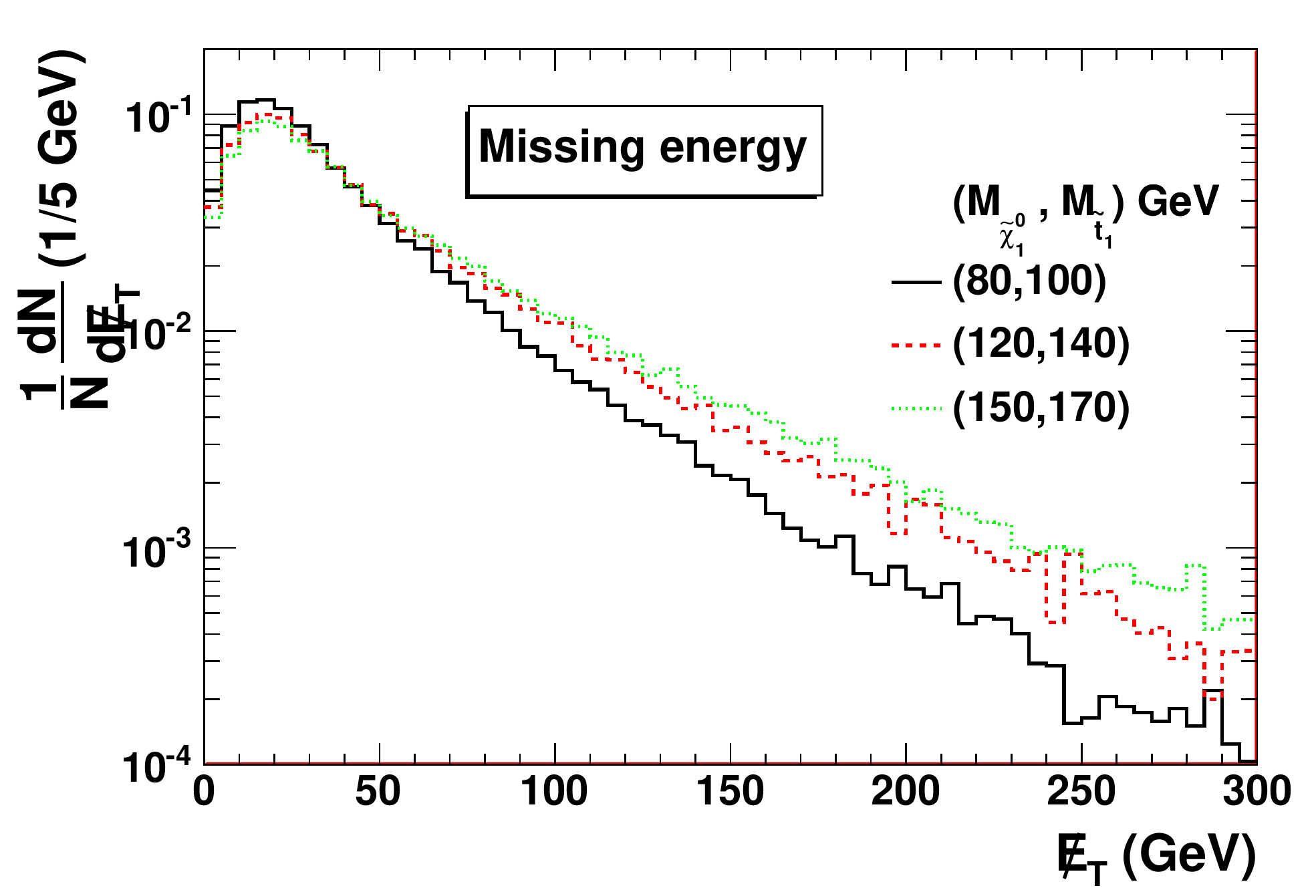}
\caption{The normalized $p_T$ distribution of the leading jet (left panel) and the missing transverse energy distribution (right panel) for increasing stop masses with fixed $M_{\tilde{t}_1}-M_{\tilde{\chi}_1^0}=20$ GeV. The normalized $p_T$ distribution of the second hardest jet (blue dot-dashed) is also shown in the left panel.}
\label{jtt}
\end{figure}

We first consider stop pair production and the same process with one additional jet at the matrix element level, assuming the gluino also decouples:
\begin{eqnarray}
pp\to \tilde{t}_1\tilde{t}_1^\ast+ j\tilde{t}_1\tilde{t}_1^\ast,
\label{pp1}
\end{eqnarray}
with a 100\% branching ratio for $\tilde{t}_1\to c\tilde{\chi}_1^0$, and with $M_{\tilde{\chi}_1^0}+M_c<M_{\tilde{t}_1}<M_{\tilde{\chi}_1^0}+M_b+M_W$ as stated before.
In Fig.~\ref{jtt} we show the normalized $p_T$ distribution of the leading jet (left panel) and the missing transverse energy distribution (right panel) for varying stop masses, with fixed $M_{\tilde{t}_1}-M_{\tilde{\chi}_1^0}=20$ GeV. One can see that the $p_T$ and missing energy distributions both peak around 20 GeV, no matter which stop and neutralino mass configuration is taken. This is because
the main source of jets here are the charms coming from stop decay which are forced to be rather soft by the assumption that $M_{\tilde{t}_1}-M_{\tilde{\chi}_1^0}=20$ GeV. Also, as the stop mass increases, more events contribute to relatively harder $p_T$ and missing energy in both distributions because the additional jet recoils against the two associated stops in the transverse direction to the beams, and thus its $p_T$ becomes harder corresponding to the increased stop mass.

\begin{figure}[tb]
\centering
\includegraphics[width=0.49\textwidth]{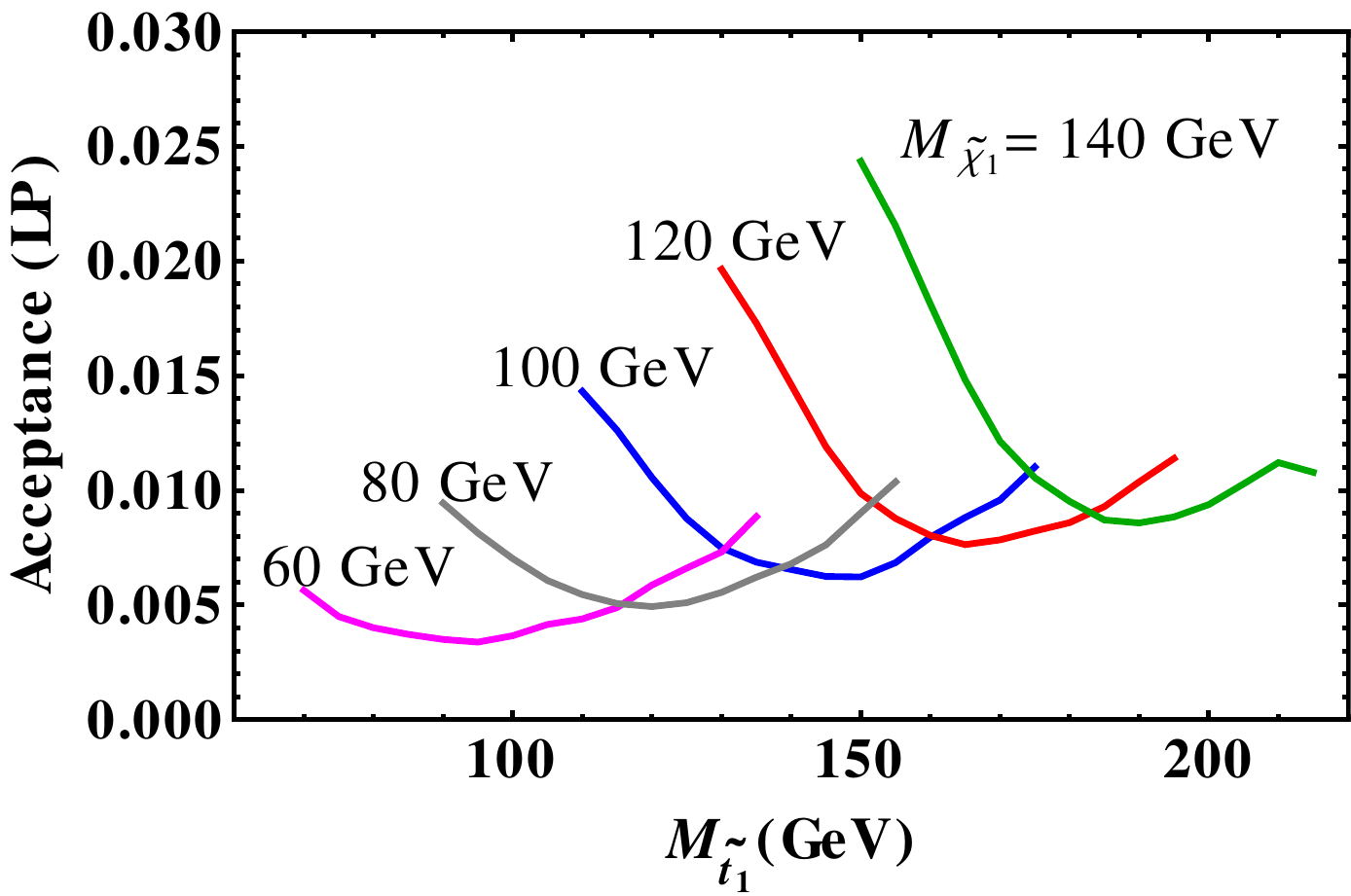}
\includegraphics[width=0.49\textwidth]{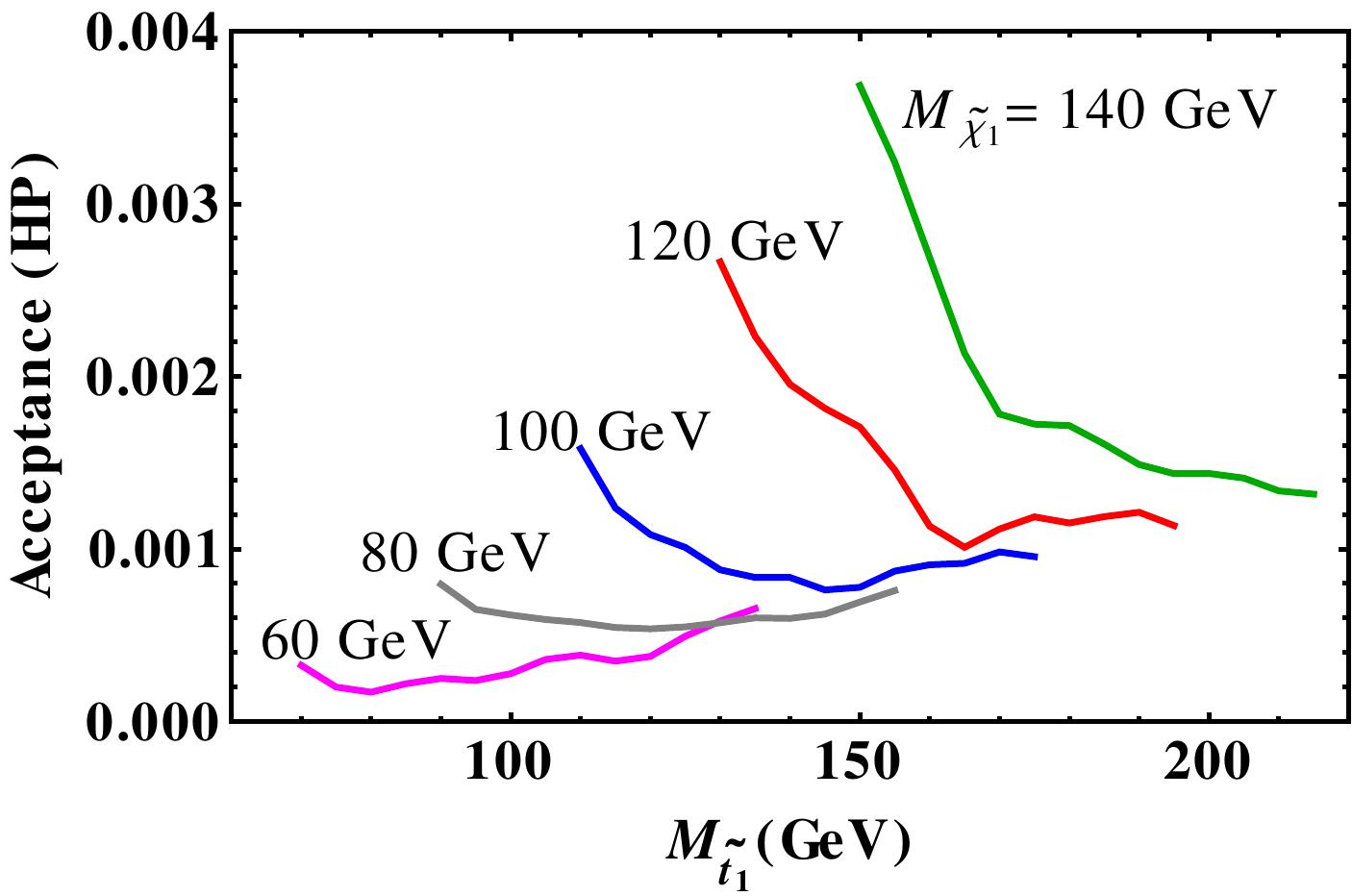}\\
\includegraphics[width=0.49\textwidth]{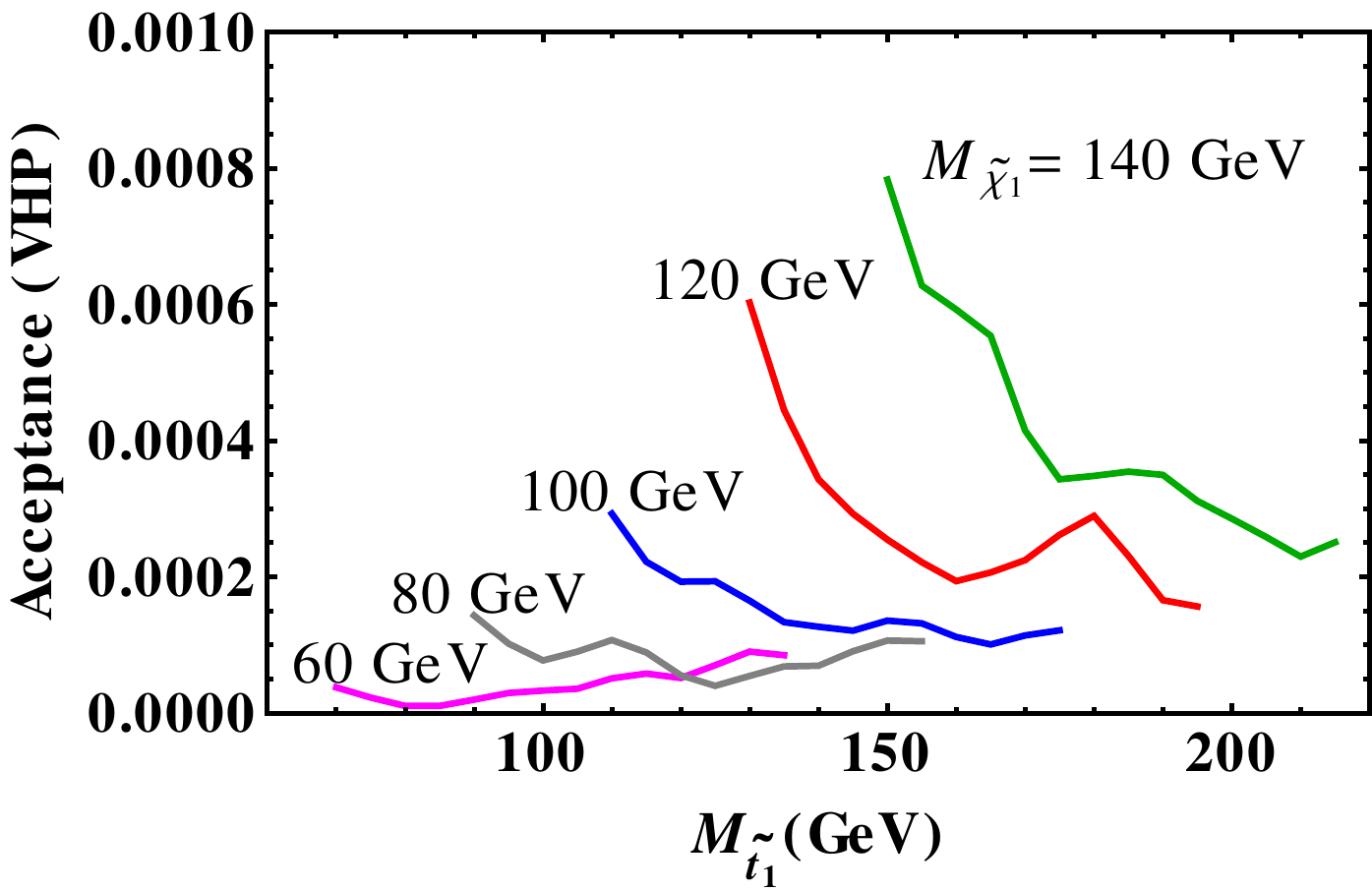}
\caption{The acceptance vs. $M_{\tilde{t}_1}$ with different masses of $\tilde{\chi}_1^0$ for the three monojet search channels LP (top left), HP (top right) and VHP (bottom).}
\label{acc}
\end{figure}

After applying the selection requirements used at the LHC, in particular for the monojet search, in Fig.~\ref{acc} we display the acceptance vs. $M_{\tilde{t}_1}$ for different $\tilde{\chi}_1^0$ masses for the three monojet search channels LP (top left), HP (top right) and VHP (bottom). As stated in the last section, the monojet signature contains one hard jet, large missing transverse energy and nothing else, and the three channels require increasing $p_T$ of the leading jet and $\cancel{E}_T$ from LP to HP to VHP. Thus, the acceptances from the above three channels reduce in the same order. One can see that for each search channel the monojet search is more sensitive to the region of small mass difference between the NLSP stop and LSP neutralino, namely $M_{\tilde{t}_1}-M_{\tilde{\chi}_1^0}\lesssim 20$ GeV for $M_{\tilde{\chi}_1^0}\gtrsim 100$ GeV, with relatively large acceptance which is sharply enhanced for heavier mass values with $M_{\tilde{\chi}_1^0}\gtrsim 120$ GeV and $M_{\tilde{t}_1}\gtrsim 130$ GeV. The features most responsible for increasing the signal acceptance in these regions are a harder $p_T$ of the additional jet and jets from the stop decay. These features respectively correspond to an increased stop mass and a lowered mass difference  $M_{\tilde{t}_1}-M_{\tilde{\chi}_1^0}$. Another consequence is that, for a fixed mass difference $M_{\tilde{t}_1}-M_{\tilde{\chi}_1^0}$, the acceptance increases with  $M_{\tilde{\chi}_1^0}$ and $M_{\tilde{t}_1}$.

\begin{figure}[tb]
\centering
\includegraphics[width=0.7\textwidth]{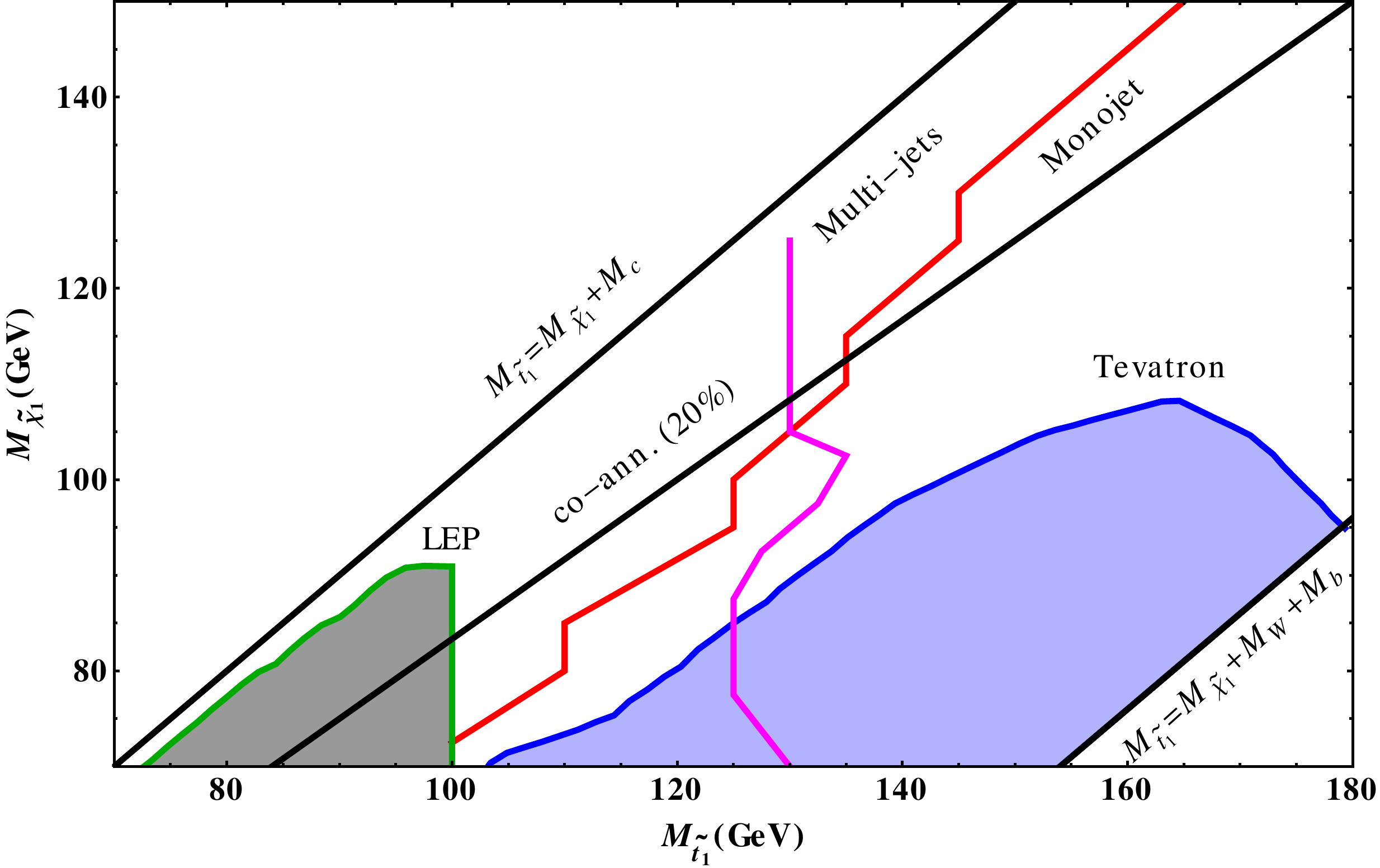}
\caption{The region excluded by LHC search for multi-jets and monojet in the $M_{\tilde{\chi}_1^0}-M_{\tilde{t}_1}$ plane. The regions excluded by LEP and Tevatron are also shown. The kinematic bounds of $\tilde{t}_1\to c\tilde{\chi}_1^0$ and $\tilde{t}_1\to bW\tilde{\chi}_1^0$ and coannihilation requirement of NLSP stop and LSP neutralino ${M_{\tilde{t}_1}-M_{\tilde{\chi}_1^0}\over M_{\tilde{\chi}_1^0}}=20\%$ are also shown for reference.}
\label{cons-stop}
\end{figure}

Based on the realization of above kinematics and LHC selection requirements and search limits on multi-jets and monojet, we make exclusion contour plot in the $M_{\tilde{\chi}_1^0}-M_{\tilde{t}_1}$ plane in Fig.~\ref{cons-stop}. Note that the Tevatron bound does not probe the region with $M_{\tilde{t}_1}-M_{\tilde{\chi}_1^0}\lesssim 40$ GeV above the LEP limit $M_{\tilde{t}_1}>100$ GeV, let alone the coannihilation region denoted in Fig.~\ref{cons-stop}. After applying the relevant selection cuts mentioned in the last section, one can see that the excluded region from the monojet search limit can reach 160 GeV for the NLSP stop mass. The 20\% coannihilation region with $M_{\tilde{t}_1}\lesssim 140$ GeV is totally ruled out. The monojet search at the LHC leaves a significant amount of unconstrained space for large $M_{\tilde{t}_1}-M_{\tilde{\chi}_1^0}$. It is consistent with the acceptance features in Fig.~\ref{acc} that for $M_{\tilde{t}_1}\gtrsim 130$ GeV, only the region with $M_{\tilde{t}_1}-M_{\tilde{\chi}_1^0}\lesssim 20$ GeV has sizable acceptance. Although having smaller acceptance, lower values of $M_{\tilde{t}_1}$ correspond to higher production cross sections for stop pair, which provide more events with a possibility of passing the selection cuts. Indeed, the region of smaller stop masses, namely $M_{\tilde{t}_1}\lesssim 130$ GeV, is excluded with even larger $M_{\tilde{t}_1}-M_{\tilde{\chi}_1^0}$ ($\sim 30$ GeV) values.

With the more stringent cuts for the subleading jets and effective mass, the search for multiple energetic jets can exclude the region of small stop masses with $M_{\tilde{t}_1}\lesssim 130$ GeV, which overlaps somewhat with the constrained region from monojet search, but extends the region to a larger mass difference, namely $M_{\tilde{t}_1}-M_{\tilde{\chi}_1^0}\gtrsim 40$ GeV. This is because the contribution of the additional hard jet provides larger missing transverse energy as well as more events containing harder jets from the stop decay, induced by the larger mass difference which can pass the multi-jets selection cuts.

\subsection{NLSP stop from heavy gluino decay}

\subsubsection{Constraints from multi-jets and monojet searches}

As previously mentioned, the NLSP stop can also be generated from heavier gluino decay and this mode introduces important search objects, like b-jets and same-sign dileptons. In this section we include gluino pair production in the processes
\begin{eqnarray}
pp\to \tilde{t}_1\tilde{t}_1^\ast+ j\tilde{t}_1\tilde{t}_1^\ast+\tilde{g}\tilde{g},
\label{pp2}
\end{eqnarray}
with the gluino decaying exclusively into a top quark and stop, namely $\tilde{g}\to t\tilde{t}_1^\ast+\bar{t}\tilde{t}_1$, with $M_{\tilde{g}}>M_{\tilde{t}}+M_t$. Also, we fix the approximate mass degeneracy of NLSP stop and LSP neutralino as ${M_{\tilde{t}_1}-M_{\tilde{\chi}_1^0}\over M_{\tilde{\chi}_1^0}}=20\%$ to remove the unknown mass parameter $M_{\tilde{\chi}_1^0}$. The total cross sections for gluino and stop pair productions are shown in Fig.~\ref{cs}. One can see that for $M_{\tilde{t}_1}<200$ GeV, the cross section for stop pair production is larger than the gluino pair production, with the gluino heavy enough to decay into an on-shell stop plus top quark. For a given integrated luminosity, most events come from stop pair production for suitably small stop mass.

\begin{figure}[tb]
\centering
\includegraphics[width=0.49\textwidth]{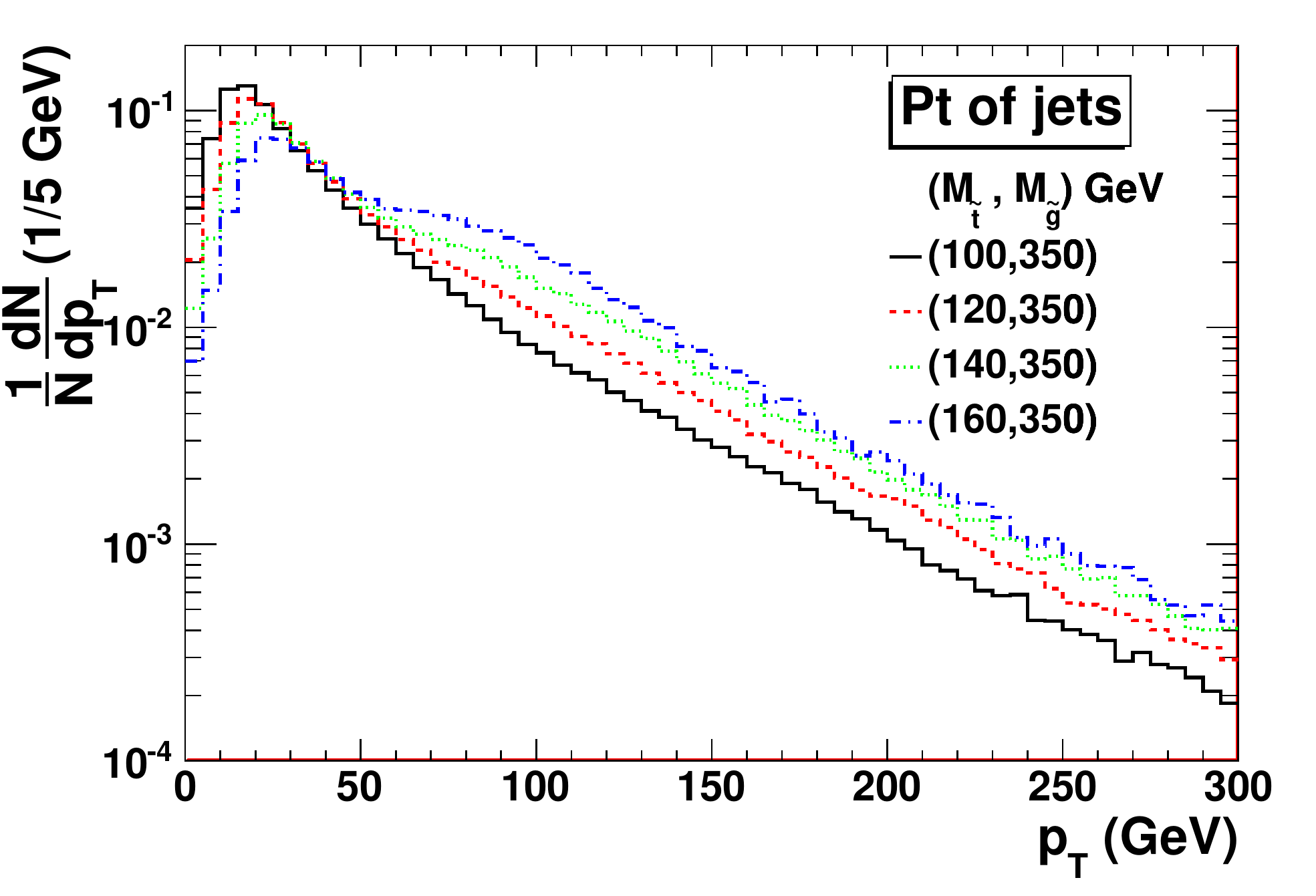}
\includegraphics[width=0.49\textwidth]{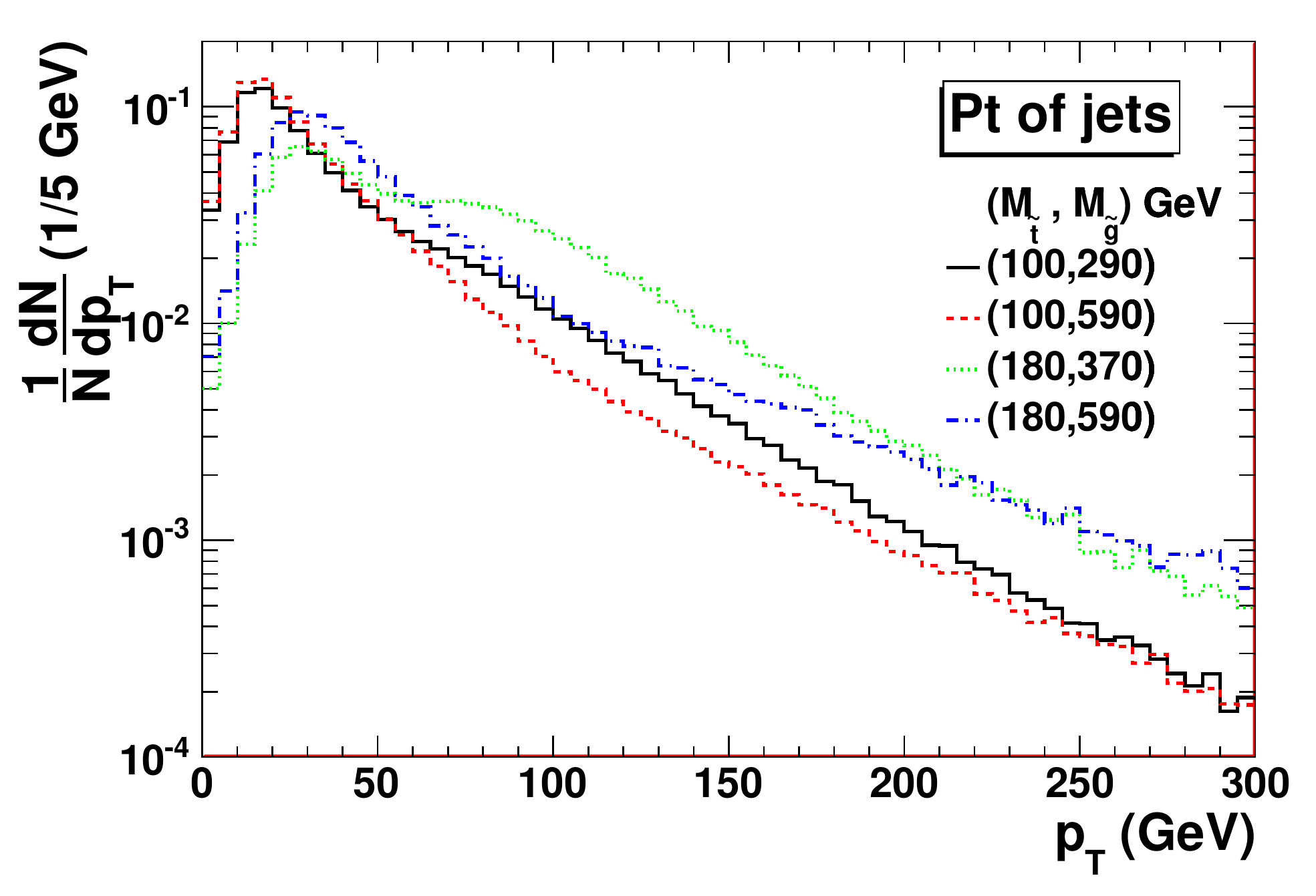}
\caption{Normalized $p_T$ distribution of leading jet for varying stop masses with fixed gluino mass (left panel), and some configurations of small and large masses of gluino and stop (right panel).
}
\label{pt}
\end{figure}

\begin{figure}[tb]
\centering
\includegraphics[width=0.49\textwidth]{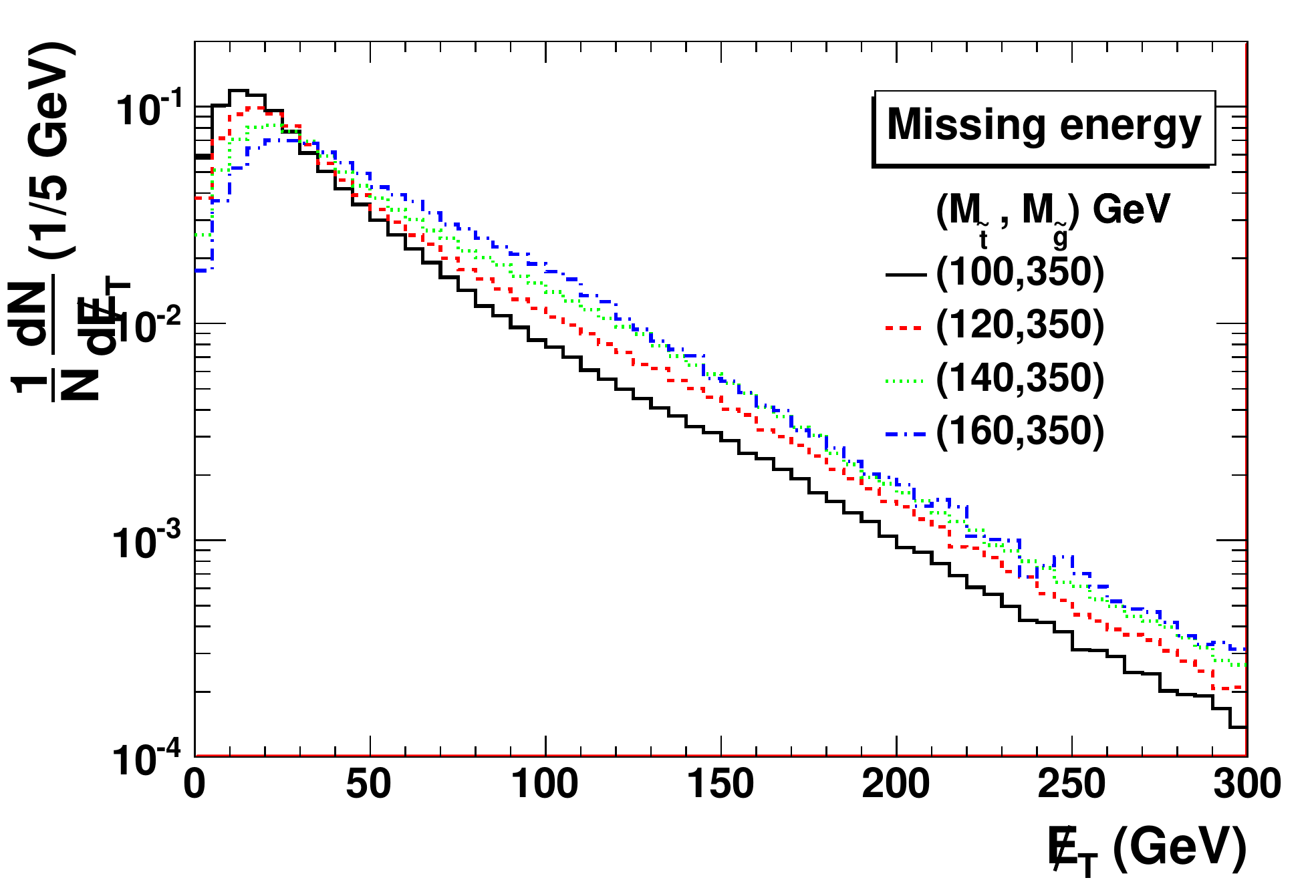}
\includegraphics[width=0.49\textwidth]{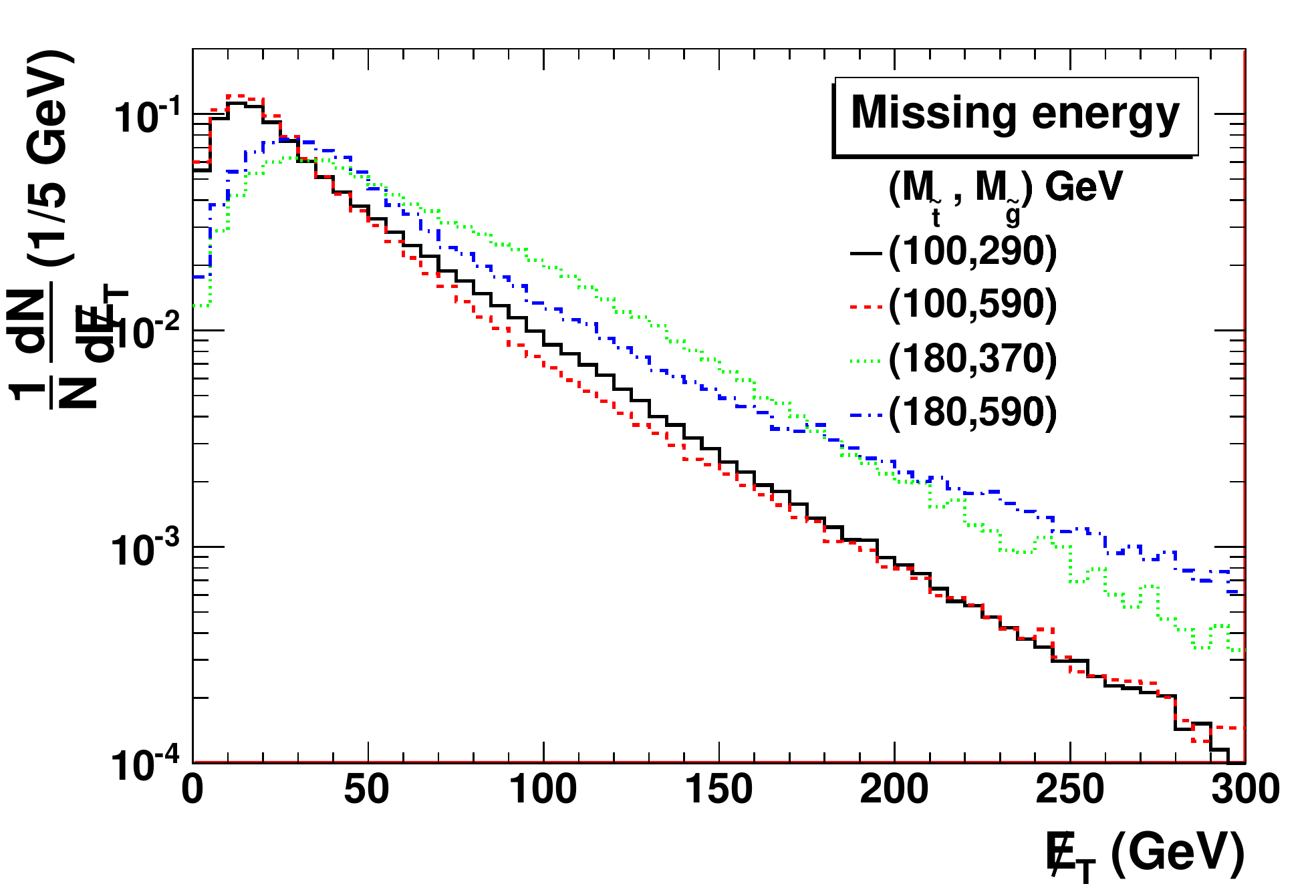}
\caption{Normalized missing energy distribution for varying stop masses with fixed gluino mass (left panel), and some configurations of small and large masses of gluino and stop (right panel).}
\label{met}
\end{figure}

\begin{figure}[tb]
\centering
\includegraphics[width=0.49\textwidth]{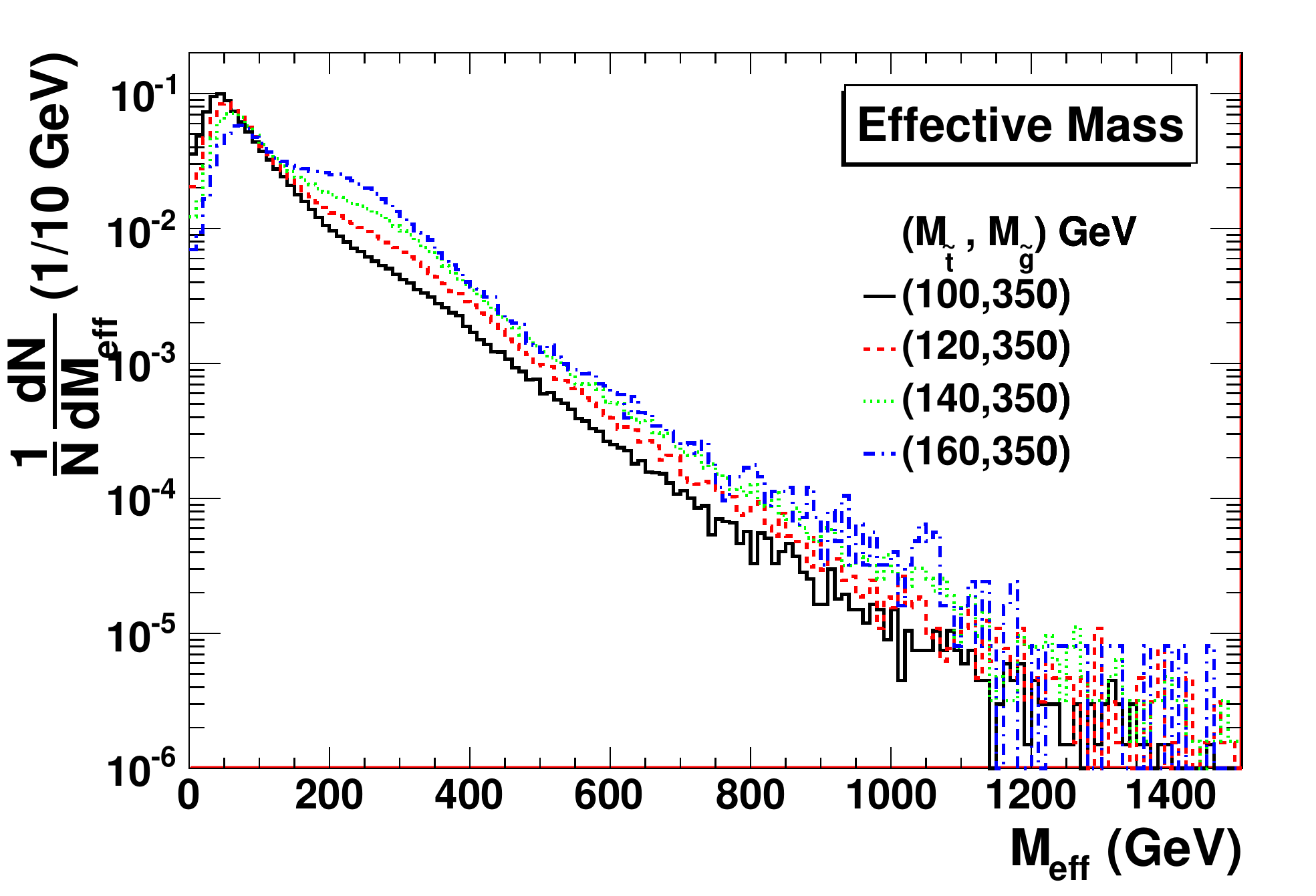}
\includegraphics[width=0.49\textwidth]{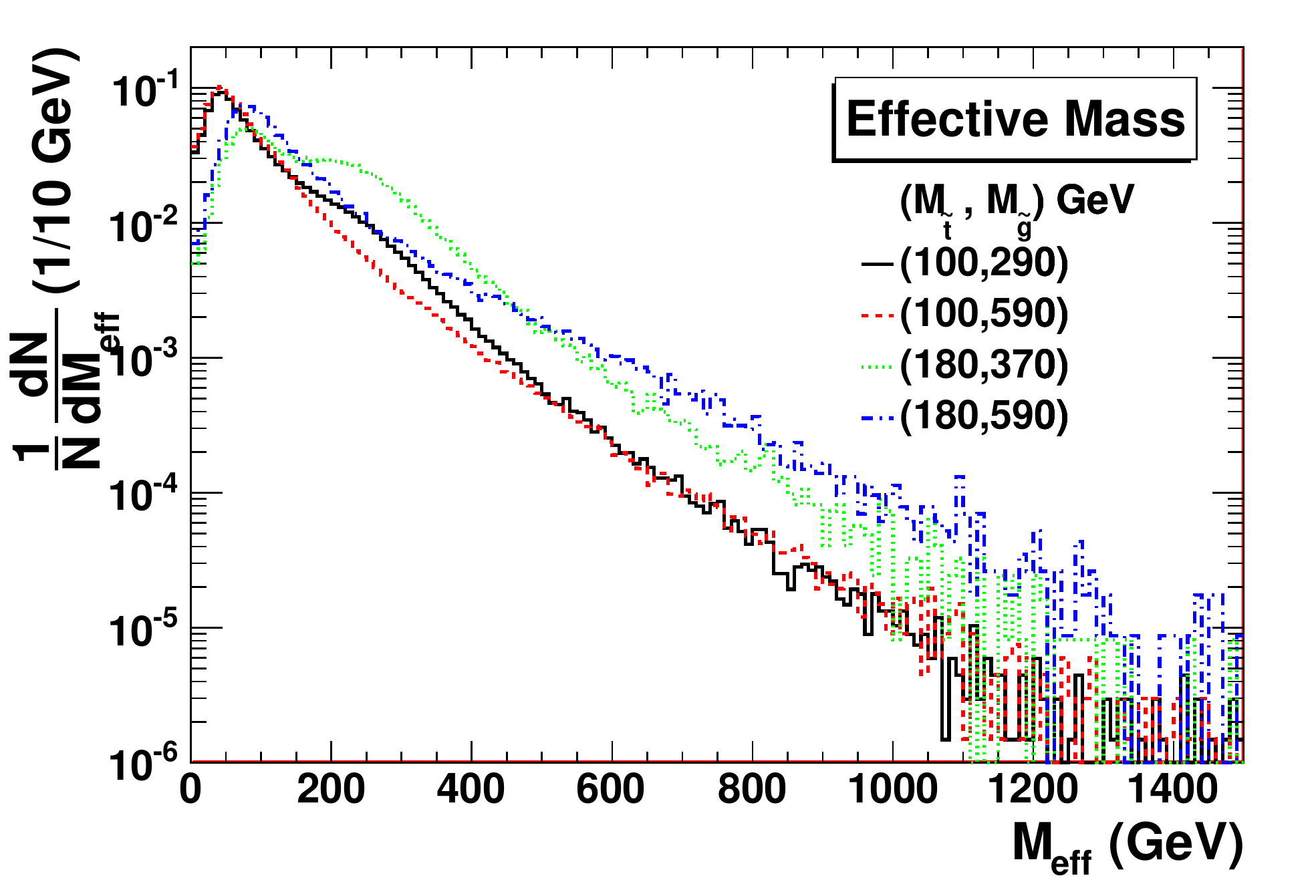}
\caption{Normalized effective mass distribution for varying stop masses with fixed gluino mass (left panel), and some configurations of small and large masses of gluino and stop (right panel).}
\label{meff}
\end{figure}

In Figs.~\ref{pt}, \ref{met} and \ref{meff} we show the normalized distributions of the leading jet $p_T$, missing transverse momentum and effective mass respectively for varying stop masses with fixed gluino mass (left panel), and some configurations with small and large masses of gluino and stop (right panel).
Because the charm jets from stop decay are the dominant source of jets,
one can see that the jet $p_T$ (missing energy and effective mass) in both plots is generally soft. But the mass difference $M_{\tilde{t}_1}-M_{\tilde{\chi}_1^0}$ increases along with the increased stop mass due to the fixed ratio ${M_{\tilde{t}_1}-M_{\tilde{\chi}_1^0}\over M_{\tilde{\chi}_1^0}}$. Thus, as shown in the left panel of Fig.~\ref{pt} (\ref{met} and \ref{meff}), the leading jet (missing energy and effective mass) becomes harder at the peak for the heavier stop production followed by stop decay into harder jets. Also, the production of heavier gluino provides events with harder products from gluino decay, especially for a heavier stop with reduced stop production cross section, for instance $p_T^{peak}(j)\sim 90$ GeV, $\cancel{E}_T^{peak}\sim 100$ GeV and $m_{eff}^{peak}\sim 250$ GeV for mass configuration $M_{\tilde{t}_1}=160$ GeV, $M_{\tilde{g}}=350$ GeV. In the right panel of Fig.~\ref{pt} (\ref{met} and \ref{meff}) one can further see the complication due to the involvement of the gluino. For the same stop mass, relatively heavier gluinos give broader distribution of jet $p_T$ (missing energy and effective mass), especially for heavier stop. This is because as the stop mass increases, the events from gluino pair production are close to those from stop pair production and the events with more energetic jets from gluino followed by top quark decay show up in the distributions.

As stated before, for signal regions S1-S4 the ATLAS analysis requires energetic jets and larger missing energy and effective mass. Consequently, most of the stop and gluino pair events with small $M_{\tilde{g}}-(M_{\tilde{t}_1}+M_t)$ would not be able to pass the relevant selection cuts. Only the region with small stop and gluino masses can be constrained due to the relatively large total cross sections. For relatively small stop masses, the stop pair events dominate due to a total cross section that is at least four times greater than that for the gluino pair production. This is followed by gluino decay into the stop with the same masses, as seen in Fig.~\ref{cs}. The relevant kinematics shown in the left panels of Figs.~\ref{pt}, \ref{met} and \ref{meff} also prohibits gluino events passing the selection cuts. So the constrained region from S1-S4 would be independent of the gluino mass for small values of stop mass. As the stop mass increases, for fixed gluino mass shown in the left panel of Fig.~\ref{pt} the enhancement of jet $p_T$ (and missing energy and effective mass) is in principle counteracted by the quickly decreasing total cross section for stop pair production. Also, the decreased gap between small $M_{\tilde{g}}$ and $M_{\tilde{t}_1}+M_t$ prevents the production of hard objects from gluino decay. Therefore, the region of relative heavier stop and extremely light or heavy gluino would evade the search bound. 
Also, because the ATLAS search for monojet plus missing energy requires only one hard jet, the relevant constraints are only sensitive to events of stop pair production associated with initial state radiation. The heavy gluino events essentially produce a boosted top quark followed by hard decay products.

\begin{figure}[tb]
\centering
\includegraphics[width=0.6\textwidth]{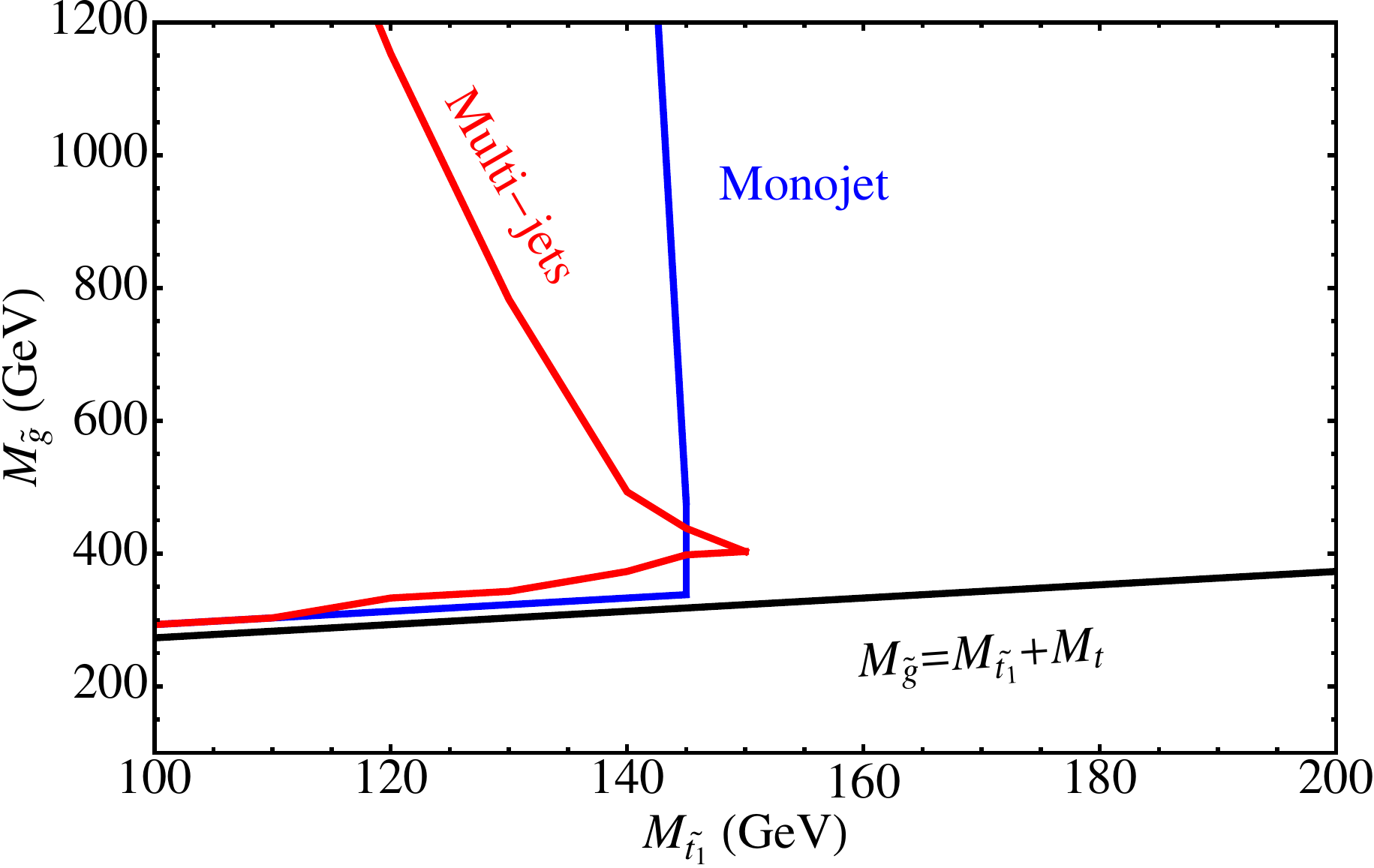}
\caption{The constrained regions in the $M_{\tilde{g}}-M_{\tilde{t}_1}$ plane using LHC data of multi-jets and monojet searches, assuming ${M_{\tilde{t}_1}-M_{\tilde{\chi}_1^0}\over M_{\tilde{\chi}_1^0}}=20\%$. The solid line represents $M_{\tilde{g}}=M_{\tilde{t}_1}+M_t$ with $M_t=173$ GeV.}
\label{cons1}
\end{figure}

In Fig.~\ref{cons1} we display in the $M_{\tilde{g}}-M_{\tilde{t}_1}$ plane the constrained region of NLSP stop scenario from LHC searches for multi-jets and monojet signals. One can see that the lower limit on the stop mass is correlated with the gluino mass from multi-jets signal regions as discussed before. The maximally ruled out stop mass is 150 GeV with a relatively low gluino mass $M_{\tilde{g}}\sim 400$ GeV. As the stop mass decreases, the constrained region favors heavier gluino masses, namely $M_{\tilde{g}}\lesssim 1$ TeV. For extremely light stop $M_{\tilde{t}_1}\lesssim 130$ GeV, the region does not depend on the gluino mass. The upper limit on the constrained stop mass is about 140 GeV from monojet signal regions, independently of the gluino mass.

\subsubsection{Constraints from b-jets and same-sign dileptons searches}

\begin{figure}[tb]
\centering
\includegraphics[width=0.49\textwidth]{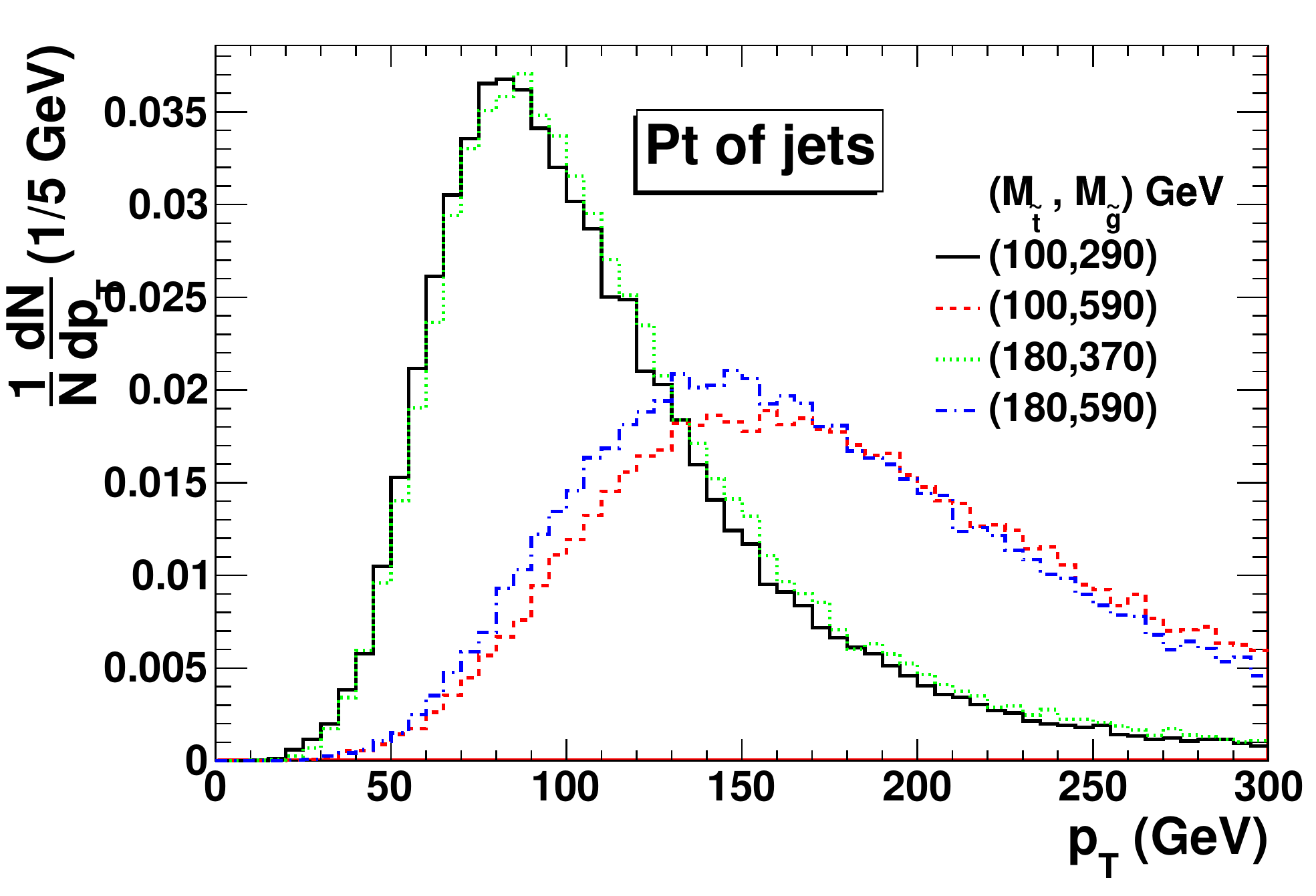}
\includegraphics[width=0.49\textwidth]{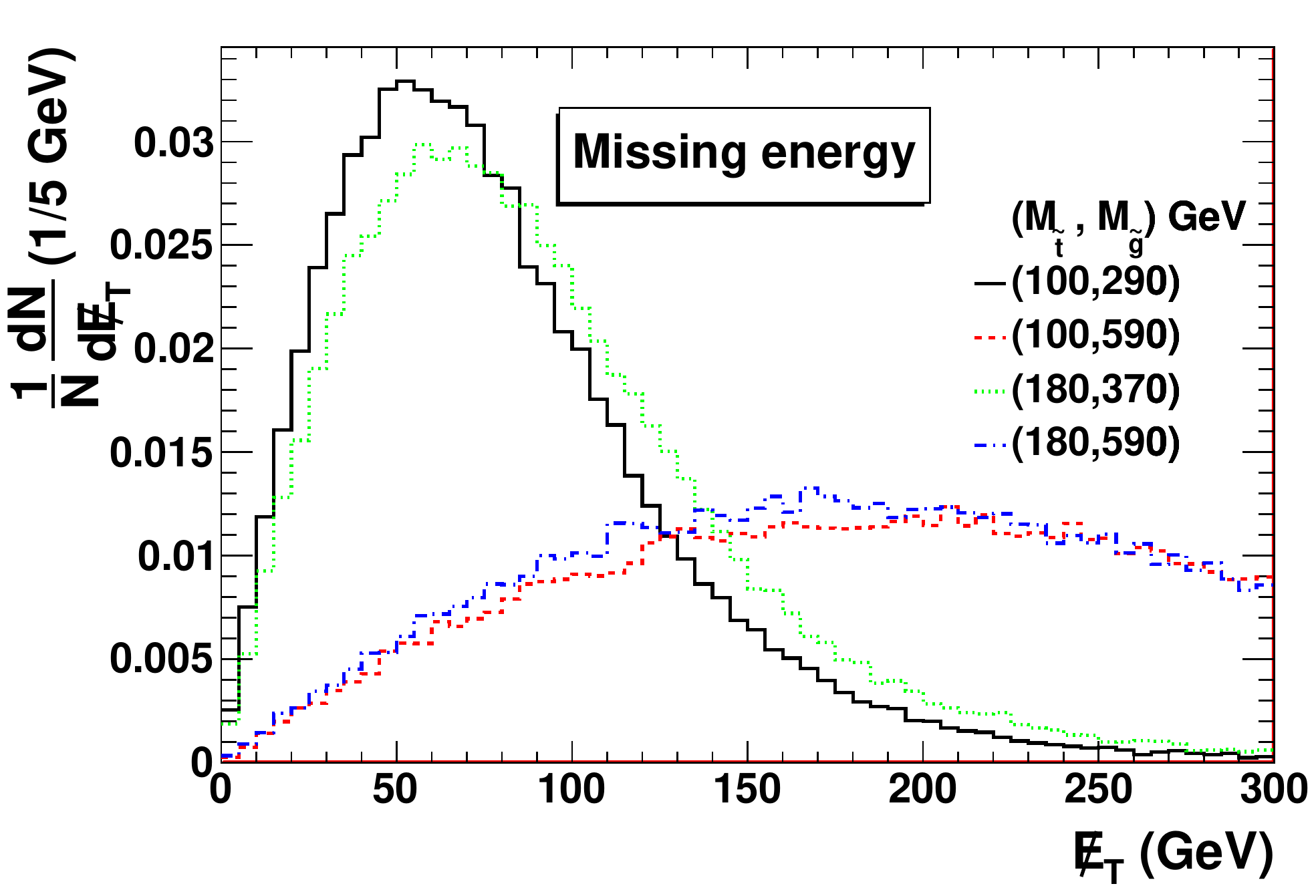}
\includegraphics[width=0.49\textwidth]{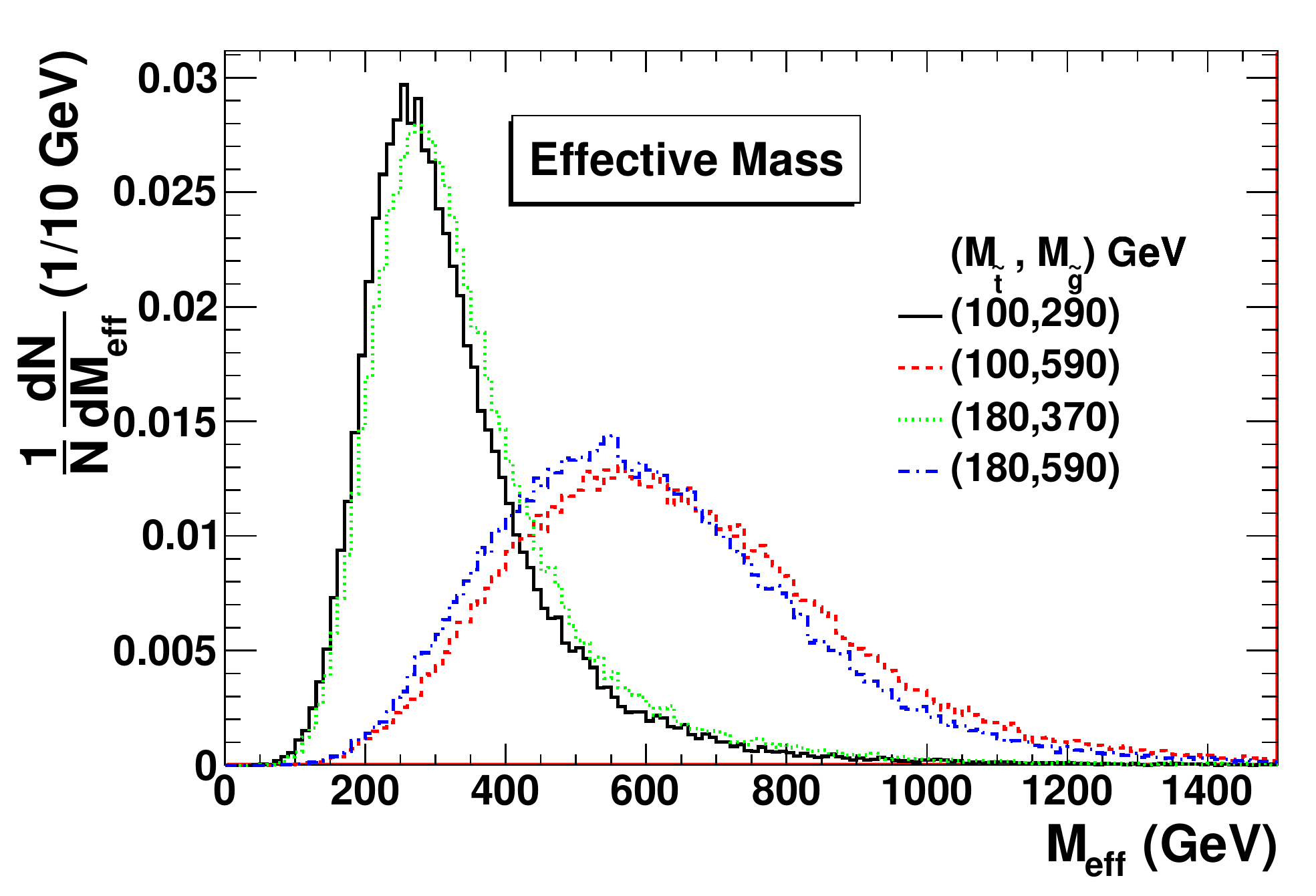}
\caption{Normalized $p_T$ of leading jet (top left), missing energy (top right) and effective mass (bottom) distributions in terms of
final states containing at least one b-jet for some configurations of small and large masses of
gluino and stop.}
\label{b}
\end{figure}

Because the gluino production discussed above is followed by gluino decay into a top quark which, in turn, essentially generates b-jets in the final states, the ATLAS searches for b-jets with or without leptons apply to the stop production events from gluino decay. The requirement of tagging b-jets would eliminate significantly events of stop pair production, so the relevant selection cuts can impose constraints on correlated gluino and stop masses in terms of stop production from gluino decay. Fig.~\ref{b} shows the normalized $p_T$ of leading jet, missing energy and effective mass distributions in terms of final states containing at least one b-jet for some configurations of small and large values of gluino and stop masses. One can see that such events generally have much more energetic jets and harder $\cancel{E}_T$ and $m_{eff}$ than those dominated by stop pair production. Also, a greater mass difference $M_{\tilde{g}}-M_{\tilde{t}_1}$ produces harder objects no matter how heavy the gluino is, and for fixed $M_{\tilde{g}}-M_{\tilde{t}_1}$, the behavior of kinematic variables are very similar as shown for the mass configurations $M_{\tilde{t}_1}=100 \ {\rm GeV}, M_{\tilde{g}}=290$ GeV and $M_{\tilde{t}_1}=180 \ {\rm GeV}, M_{\tilde{g}}=370$ GeV. For a fixed gluino mass, the kinematics in Fig.~\ref{b} become softer for increased stop masses (because of smaller mass difference $M_{\tilde{g}}-M_{\tilde{t}_1}$), and thus softer products from top quark decay. Therefore, we expect a lower limit on the stop mass in the constrained region from b-jets search.

The same-sign dilepton search also only picks up gluino pair production followed by same-sign top quarks, both of which undergo leptonic decay. So it is sensitive to both gluino and stop masses as well. Because the relevant selection cuts are much less stringent than those in b-jets signal search and the requirement of same-sign dileptons helps to significantly reduce the SM backgrounds, we expect the constrained region to be much broader than that from b-jets search, although there is suppression arising from the top branching ratio to leptons.

\begin{figure}[tb]
\centering
\includegraphics[width=0.6\textwidth]{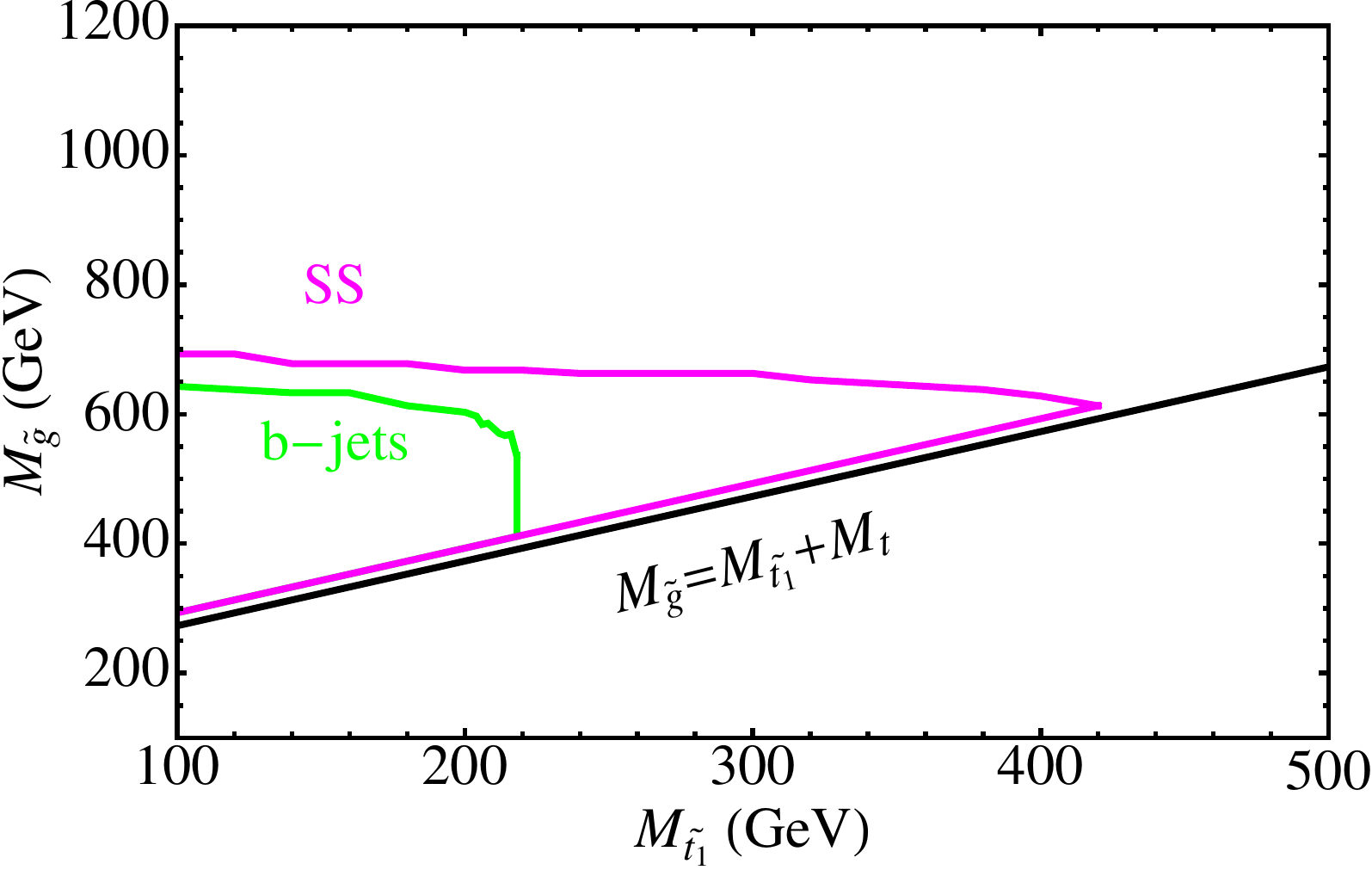}
\caption{The constrained regions in the $M_{\tilde{g}}-M_{\tilde{t}_1}$ plane using LHC data of b-jets and same-sign dileptons (SS) searches, assuming ${M_{\tilde{t}_1}-M_{\tilde{\chi}_1^0}\over M_{\tilde{\chi}_1^0}}=20\%$. The solid line represents $M_{\tilde{g}}=M_{\tilde{t}_1}+M_t$ with $M_t=173$ GeV.}
\label{cons2}
\end{figure}

In Fig.~\ref{cons2} we display in the $M_{\tilde{g}}-M_{\tilde{t}_1}$ plane the excluded region of NLSP stop scenario from heavier gluino decay by LHC data on b-jets and same-sign dileptons (SS) searches. One can see that the LHC data imposes a lower limit of about 600 GeV and 700 GeV on the gluino mass in this scenario from b-jets and same-sign dileptons searches respectively. The lower limit on the stop mass from b-jets search is around 220 GeV and the limit from same-sign dileptons is 400 GeV as expected.

Note that the ATLAS and CMS experiments have also presented their analysis results for events containing jets and one lepton. We do not list this channel here because in the process in Eq.~(\ref{pp1}), there is no lepton in the final state. Besides, in the process in Eq.~(\ref{pp2}) with the gluino decaying exclusively into a top quark and stop, the top quark generates b-jets in the final state. For this scenario we have applied the analysis results for b-jets with or without leptons.

\section{Summary}
We study the NLSP stop scenario which arises from implementing $b-\tau$ Yukawa coupling unification in the CMSSM framework. We consider a simplified spectrum with only LSP neutralino mass $M_{\tilde{\chi}_1^0}$, NLSP stop mass $M_{\tilde{t}_1}$ and heavier gluino mass $M_{\tilde{g}}$ at low energy. The light stops are produced in pairs in association with a hard jet, or as decay products from heavier gluino production, namely $\tilde{g}\to t\tilde{t}_1^\ast+\bar{t}\tilde{t}_1$ with 100\% branching fraction. The two-body mode $\tilde{t}_1\to c\tilde{\chi}_1^0$ is assumed to be the unique stop decay channel.

We have employed the ATLAS and CMS searches, corresponding to 1 fb$^{-1}$ of integrated luminosity, to impose constraints on this scenario. In the neutralino-stop coannihilation region that we are primarily interested in, we were able to show that NLSP stop masses below around 140-160 GeV are essentially ruled out or strongly constrained. We also obtain a lower bound in this scenario of 600 GeV (700 GeV) on the gluino mass from b-jets (same-sign dileptons) searches.

\subsection*{Acknowledgment}
We would like to thank M. Adeel Ajaib, Yevgeny Kats, Jing Shao and Daniel Feldman for useful discussions. This work is supported by the DOE under grant No. DE-FG02-91ER40626. This work used the Extreme Science and Engineering Discovery Environment (XSEDE), which is supported by National Science Foundation grant number OCI-1053575.



\end{document}